\documentclass[twocolumn, trackchanges,times,twocolappendix]{aastex631}

\begin{document}

\title[Kinematics of PSBs]{Spatially Resolved Kinematics of Post-starburst Galaxies}

\shorttitle{Kinematics of Post-starbursts}
\shortauthors{Pardasani \& French}

\author[0000-0002-7496-9768]{Bhavya Pardasani}\thanks{Email: bhavya.pardasani@yale.edu}
\affiliation{Department of Astronomy, Yale University, Kline Biology Tower, 219 Prospect St, New Haven, CT 06511, USA}
\affiliation{Department of Astronomy, University of Illinois at Urbana-Champaign, 1002 W Green St, Urbana, IL 61801, USA}

\author[0000-0002-4235-7337]{K. Decker French}
\affiliation{Department of Astronomy, University of Illinois at Urbana-Champaign, 1002 W Green St, Urbana, IL 61801, USA}

\begin{abstract}
Star-forming galaxies can reach quiescence via rapid transition through merger-triggered starbursts that consequently affect both their kinematics and star formation rates. In this work, we analyze the spatially resolved kinematics of 89 post-starburst galaxies (PSBs) with data from the Mapping Nearby Galaxies at Apache Point Observatory (MaNGA) survey and place them in context with star-forming galaxies (SFGs) and early-type galaxies (ETGs) to study the impact of merger history on galaxy kinematics. We measure the specific angular momentum to characterize them as slow or fast rotators. We find that the MaNGA PSB sample has $\sim 6\%$ slow rotators, which is less than the $\sim 14\%$ slow rotators in the ATLAS$^{3D}$ ETG sample and $\sim 20\%$ slow rotators in the MaNGA ETGs, and more than the $\sim 3.5\%$ slow rotators in the MaNGA SFGs. This implies that for PSBs to evolve into ETGs, they must still lose some angular momentum. While ETGs with higher stellar mass tend to be slow rotators, PSBs do not follow this trend. We find significant correlations between specific angular momentum and mass-weighted age for the SFG and ETG samples, but do not see any significant trends within the short PSB phase. These results indicate that significant evolution in angular momentum must continue to take place as the galaxy ages after the PSB phase. For PSBs to evolve directly into ETGs, they must undergo dry mergers to shed excess angular momentum without causing further epochs of star formation.

\end{abstract}



\section{Introduction} 
\label{sec: intro}

Galaxies can be classified into two groups: star-forming and quiescent, which also divides galaxies into disk-like and early-type morphologies. Merger events can cause galaxies to evolve from star-forming disks to quiescent ellipticals. There is a relatively short transitional phase of $\sim 250$ Myr after a rapid quenching event known as the post-starburst phase (``E+A" or ``K+A"). Post-starburst galaxies (PSBs) reside in the green valley region between the red sequence of the early-type galaxies (ETGs) and blue cloud of the star-forming galaxies (SFGs). They lack the H$\alpha$ emission lines characteristic of SFGs that have younger and hotter O and B stars, but show strong H$\delta$ Balmer absorption, which indicates an abundance of A stars that are absent in long-quiescent ETGs \citep{1973ApJ...182..381S, 1978IAUS...77..279S, 1982ApJ...252..455S, 1983ApJ...270....7D, 1987MNRAS.229..423C, 2021PASP..133g2001F}, indicating a recent burst of star formation. The PSB phase is short and thus PSBs are rare, especially in the local Universe. However, integrated over cosmic time, $\sim40\%-100$\% of ETGs likely went through an episode of rapid quenching \citep{2011ApJ...741...77S, 2020MNRAS.494..529W}.

In order to trace the evolution of a galaxy and any mergers undergone by it, we study its kinematics. Resolved integral field unit (IFU) spectroscopy gives information about the rotational and dispersion components of the stellar velocity, which lets us quantify the specific angular momentum, as done by \citet{2007MNRAS.379..401E}, and study the relative significance of stellar rotational velocity ($v$) and stellar velocity dispersion ($\sigma$). The specific angular momentum ($\lambda_R$) of a galaxy provides information about whether a galaxy is a slow ($\lambda_R \lesssim 0.1$) or a fast ($\lambda_R \gtrsim 0.1$) rotator \citep{2007MNRAS.379..401E}. A comparison of this quantity for PSBs with other galaxies, especially ETGs, can help us connect the merger history with evolution in stellar populations.

The angular momentum content of galaxies is sensitive to their recent merger history because the geometry of the merger and merger mass ratio will affect how much angular momentum is lost or gained. Classifying galaxies kinematically, based on $\lambda_R$, can offer valuable insight into their merger history. A loss in angular momentum can happen when two similar mass galaxies merge, in a major merger, along their centers such that they decrease the angular momentum of the resulting galaxy \citep{2003ApJ...597..893N}. A gain in angular momentum could be a product of a merger between a massive galaxy colliding with a significantly lower-mass galaxy (a minor merger) with a high value of angular momentum, which leads to the resulting galaxy gaining the excess momentum \citep{2003ApJ...597..893N}. If the merger does not result in a huge influx of cold gas (a dry merger), it can lead to a loss in angular momentum without triggering further star formation.

Observations of galaxies in the immediate aftermath of a strong starburst can be used to test the origins of angular momentum content in ETGs. Major mergers are thought to play a key role in the creation of ETGs \citep{2008ApJS..175..356H}, and thus we would expect the ETGs to be slow-rotating and dispersion dominated. However, \citet{2007MNRAS.379..401E, 2011MNRAS.414..888E} found that ETGs consist of both slow and fast rotators. \citet{2007MNRAS.379..401E} suggest that dissipationless (or ``dry'') mergers, which cause angular momentum to be expelled outwards, play a key role in the evolution of slow rotators. They propose that multiple channels including dry mergers, gas accretion, and wet mergers likely play a role in the evolution of angular momentum in different classes of ETGs. A key test of how and when slow-rotator ETGs lose their angular momentum is the fraction of slow rotators in galaxies in the immediate aftermath of a merger. PSB galaxies appear to be in the middle of evolving from the starburst phase to the early-type phase, likely driven by a recent major merger between two gas-rich galaxies \citep{1996ApJ...466..104Z, 2008ApJ...688..945Y, 2009MNRAS.395..144W, 2016MNRAS.456.3032P, 2021ApJ...919..134S}, providing a snapshot into the ETG population that may result from this rapid channel of evolution. 

\citet{2001ApJ...557..150N}, using long-slit spectroscopic data, found 18 out of 20 PSBs from \citet{1996ApJ...466..104Z} to be dispersion dominated, i.e. $v/\sigma < 1$, and only six out of 20 PSBs exhibited rotation with $v>40$ km s$^{-1}$. \citet{2009MNRAS.396.1349P}, \citet{2012MNRAS.420..672S} and \citet{2013MNRAS.432.3131P} used integral field spectroscopy to analyze the kinematics of PSBs and found that the PSBs are primarily dominated by fast rotators. \citet{2012MNRAS.420.2232P} and \citet{2013MNRAS.432.3131P} used the ellipticity-dependent classification presented in \citet{2011MNRAS.414..888E} to find four slow rotators in a sample of 26 E+A galaxies at $z \sim 0.02-0.04$. They found that the slow-rotator fraction for the PSBs ($\sim 15\%-17\%$) is similar to the slow-rotator fraction in the ETGs from the ATLAS$^{3D}$ sample ($14\%$; \citet{2011MNRAS.414..888E}). Similarly, \citet{2012MNRAS.420..672S} found $\sim9\%$ slow rotators in a sample of 11 E+A galaxies at $z = 0.06-0.12$. In contrast to findings by \citet{2001ApJ...557..150N} that are more consistent with PSBs experiencing major mergers, results from \citet{2012MNRAS.420..672S} and \citet{2013MNRAS.432.3131P} suggest that their sample of PSBs probably experienced minor mergers. However, a lack of slow rotators in PSBs does not automatically discard the possibility of a major merger as even a 2:1 mass ratio merger can lead to a fast rotator \citep{2011MNRAS.414..888E} after triggering a starburst. Moreover, simulated slow rotators usually lose their angular momentum over a period of quite a few billion years via dry mergers \citep{2014MNRAS.444.3357N}. While it is clear that most PSBs are not already slow rotators, their dispersion-dominated kinematics and fraction of slow rotators are similar to ETGs within small number uncertainties. Larger samples of PSBs with spatially resolved kinematics are needed to test how the angular momentum content of quiescent galaxies evolves over time, and the role of major and minor mergers in the formation of slow-rotator ETGs.

In order to further test and compare the results obtained by earlier works, we use integral field spectroscopic data for a large sample of 89 PSBs from the Mapping Nearby Galaxies at Apache Point Observatory \citep[MaNGA;][]{2015ApJ...798....7B, 2016AJ....152..197Y, 2022ApJS..259...35A} survey, which provides high-resolution spectroscopy for the inner kinematics of 10,000 nearby galaxies. We perform similar $\lambda_R$ calculations to \citet{2007MNRAS.379..401E} for our PSB sample to test whether the PSBs are a result of major mergers, are evolving into ETGs or have undergone various minor/dry mergers. If all of the PSBs have been subjected to equal-mass major mergers, we would expect them to have more slow rotators as compared to the ETGs, with perhaps another secular channel producing fast-rotator ETGs. If wet mergers are the main mechanism to produce ETGs, with slow rotators produced in the immediate aftermath of the merger-triggered starburst, we would expect PSBs and ETGs to have a comparable fraction of slow rotators. If minor dry mergers are required to produce slow-rotator ETGs, we would expect to see fewer slow rotators among the PSB sample than in the ETG sample.

We present the data and methods in Section \ref{sec: data_and_methods}. In Section \ref{sec: calculation}, we outline the calculation and corrections performed for the specific angular momentum using the angular momentum expression given by \citet{2007MNRAS.379..401E}. We present our results from our analysis of angular momentum as a function of ellipticity, mass-weighted (MW) age, light-weighted (LW) age, and asymmetry in Sections \ref{sec: results} and \ref{sec: discussion}. Finally, we summarize our results and the implications from this work in Section \ref{sec: conclusion}.


\begin{figure*}
\centering
\epsscale{1.1}
\plotone{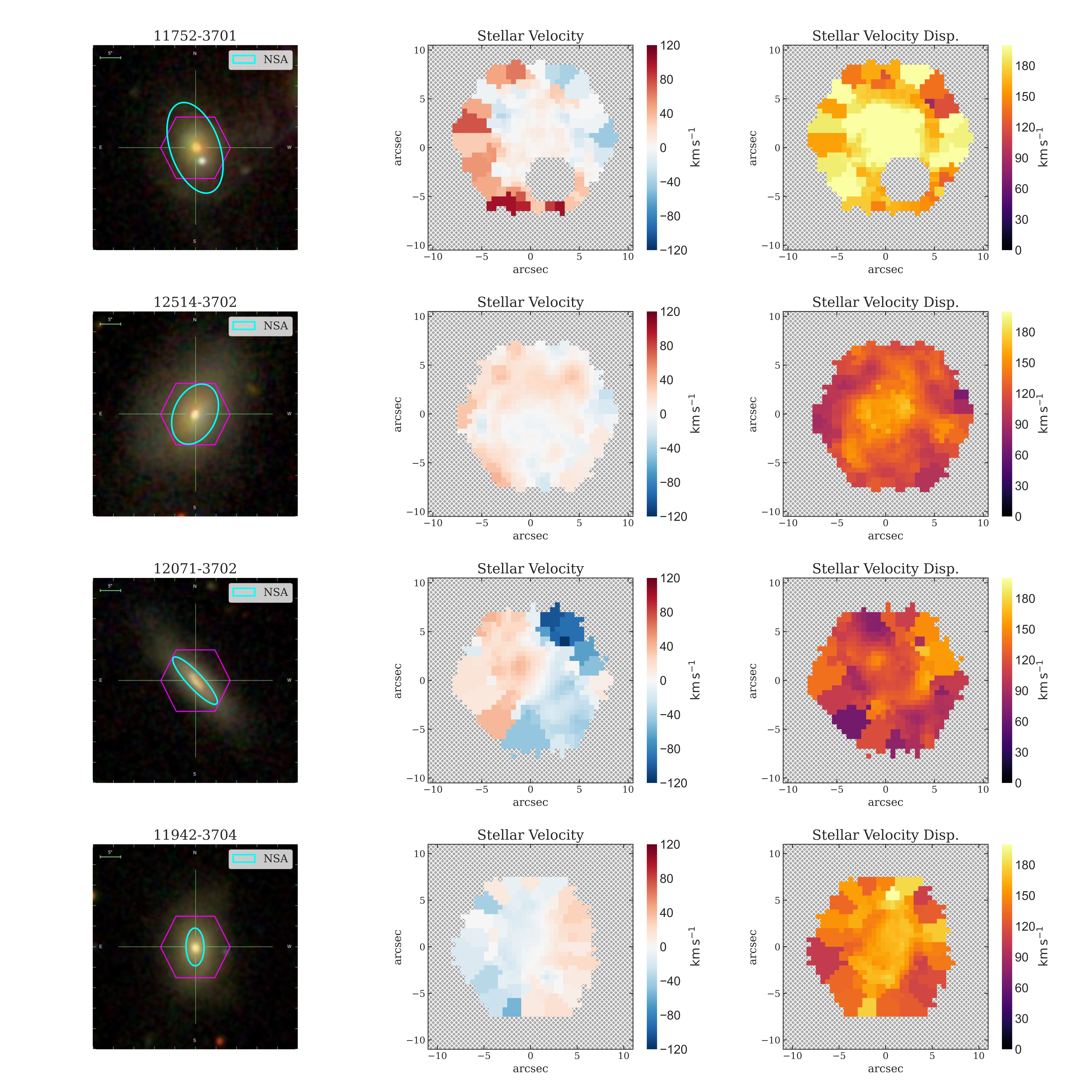}
\caption{The Marvin visualization of the MaNGA data for four of the identified slow rotators in the post-starburst galaxy sample. The left-most panel shows the SDSS \textit{gri} postage stamp image of each galaxy, with the pink hexagon representing the MaNGA field of view, and the title gives the plate-ifu that can be used to identify the galaxy in Marvin and MaNGA. The middle and rightmost panels show the corresponding stellar velocity and stellar velocity dispersion, each having a fixed color bar across the galaxies. The solid blue ellipse is created using the photometric properties from the NSA catalog.
\label{fig: slow_rotators}}
\end{figure*}

\begin{figure*}
\centering
\epsscale{1.1}
\plotone{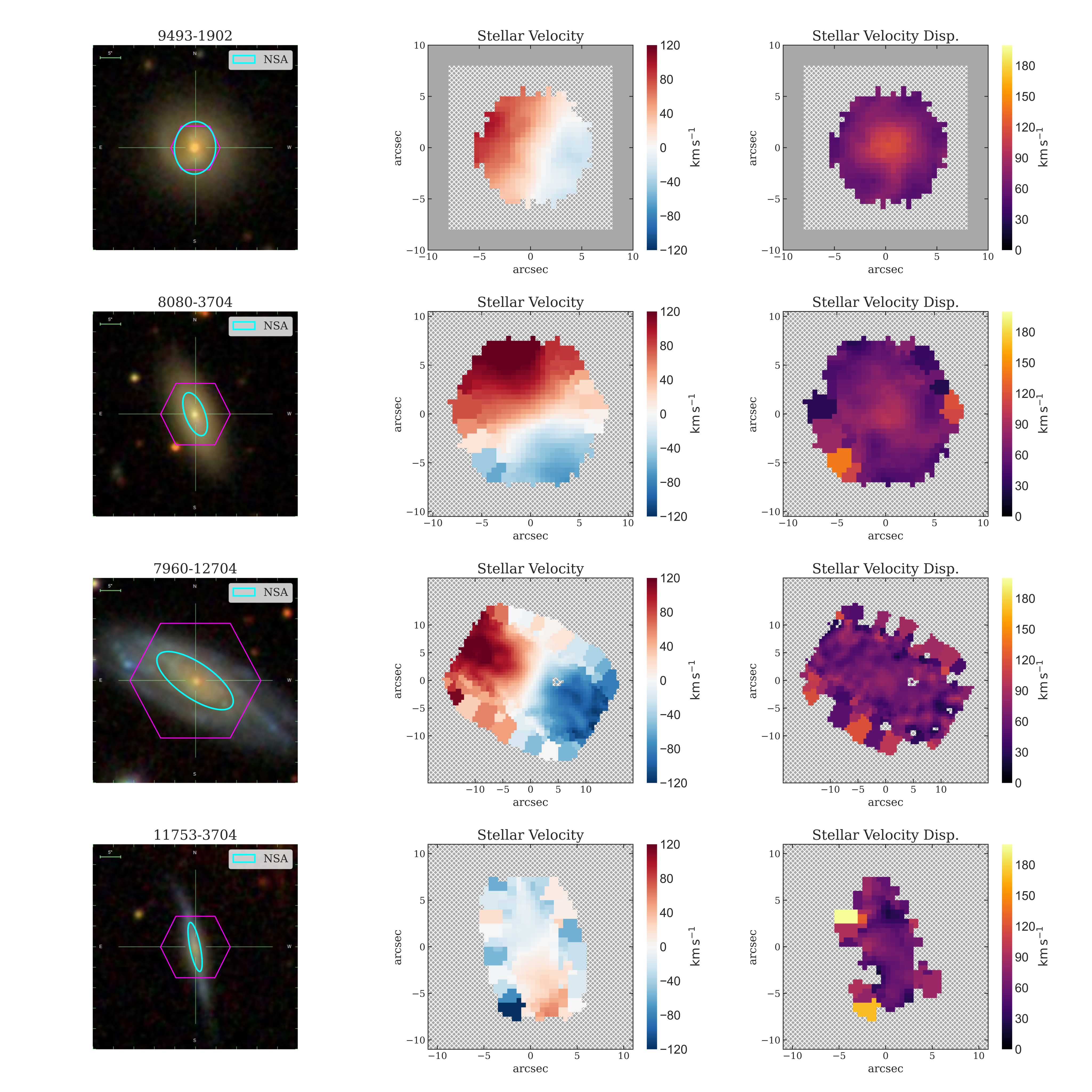}
\caption{The Marvin visualization of the MaNGA data for four of the identified fast rotators in the post-starburst galaxy sample. The left-most panel shows the SDSS \textit{gri} postage stamp image of each galaxy, with the pink hexagon representing the MaNGA field of view, and the title gives the plate-ifu that can be used to identify the galaxy in MaNGA. The middle and rightmost panels show the corresponding stellar velocity and stellar velocity dispersion, each having a fixed color bar across the galaxies. The solid blue ellipse is created using the photometric properties from the NSA catalog.
\label{fig: fast_rotators}}
\end{figure*}

\section{Data and Target Selection}
\label{sec: data_and_methods}

\subsection{Post-starburst Galaxy Selection}
\label{sec: sample_selection}

There are many methods in the literature for selecting galaxies that have experienced a recent starburst \citep[see][for a recent review]{2021PASP..133g2001F}. This work selects galaxies with a variety of post-starburst ages and active galactic nuclei (AGN) properties from a combination of three different samples of PSBs identified in the Sloan Digital Sky Survey (SDSS; \citealt{2000AJ....120.1579Y,2002AJ....124.1810S}), as used in \citet{2023ApJ...950..153F}. 

The first sample of E+A galaxies from  \citet{2018ApJ...862....2F} were identified based on H$\delta$ strength to select for a dominant A-star population and a criterion against H$\alpha$ emission to select against ongoing star formation, which is similar to methodologies employed by \citet{1996ApJ...466..104Z, 2005MNRAS.357..937G}. This stringent restriction against star formation ensures the quiescent nature of these galaxies but also excludes galaxies with strong AGN. The sample of Shocked PSBs (SPOGs; \citealt{2016ApJ...830..137A}) uses a similar selection for strong Balmer absorption, but allows for stronger emission lines, requiring galaxies to exhibit emission-line ratios inconsistent with those expected from star formation \citep{1981PASP...93....5B,2006MNRAS.372..961K}. The sample of principal component analysis (PCA)-selected PSBs from \citet{2007MNRAS.381..543W, 2009MNRAS.395..144W} also allows for stronger emission lines from either star formation or AGN. The SPOG and PCA galaxies generally skew younger and have slightly higher star formation rates as compared to the E+A galaxies \citep{2018ApJ...862....2F,2018ApJ...863...28A}. There exists some overlap among these different samples, culminating in a combined sample size of 5040 galaxies.

\subsection{MaNGA}
\label{sec: MaNGA}

The MaNGA Survey \citep{2015ApJ...798....7B, 2016AJ....152..197Y, 2022ApJS..259...35A} is one of the three surveys of SDSS-IV that provides data collected for the composition and kinematic structure of $\sim$10,000 nearby galaxies via the Sloan 2.5 m telescope at the Apache Point Observatory. It utilizes optical circular fibers to group 29 fiber bundles of various sizes to create hexagonally shaped IFUs that are used to spectroscopically observe various sections of a galaxy for a wavelength range of $360-1040$ nm and a spectral resolution of 2000 \citep{2015AJ....149...77D, 2015AJ....150...19L, 2017AJ....154...86W}. The fibres have a core diameter of 120 $\mu$m which subtends to $2\arcsec$ on the sky. The Sloan telescope has 17 IFUs of various sizes that are used to observe different galaxies simultaneously within a field of view of $3^{\circ}$ diametrically ($7$ deg$^2$). 

We compared the R.A. and decl. of the galaxies from the Data Release 17 (DR17) version of MaNGA to the $\sim$5000 galaxies from the combined post-starburst sample described in Section \ref{sec: sample_selection} to select 89 PSBs\footnote{One galaxy is dropped from the sample used by \citet{2023ApJ...950..153F} because it is not included in the MPA-JHU galSpec analysis, which we use to obtain the stellar mass. Three galaxies were dropped as they are undergoing mergers (based on their stellar velocity maps), which can significantly impact their angular momentum, thus introducing inaccuracies in kinematic analysis. This leaves us with 89 PSBs out of the 93 in \citet{2023ApJ...950..153F}.}. We used NASA-Sloan Atlas (NSA) catalog \citep{2011AJ....142...31B} version {\tt v1\_0\_1} to obtain the photometric properties like Sérsic index ({\tt SERSIC\_N}), ellipticity ($\epsilon = 1 - ${\tt SERSIC\_BA}), position angle ({\tt SERSIC\_PHI}) and effective radius ({\tt SERSIC\_TH50}) in the $r$ band, all of which were calculated using a Sérsic fit, for the MaNGA galaxies. \citet{2018MNRAS.477.4711G} calculated and compared the angular momentum for 2300 galaxies of various types from MaNGA Data Release 14 (DR14) data, and identified the slow and fast rotators in their sample. We remove the 16 PSBs we identified from their sample, as our angular momentum calculations for the PSBs are based on the MaNGA survey's DR17 version. The remaining galaxies (SFGs and ETGs) are cross-matched with the NSA catalog to reproduce their angular momentum calculation using MaNGA DR17 and NSA photometry. These MaNGA galaxies are utilized for comparisons and analysis in multiple upcoming plots. We prefer MaNGA release DR17 over DR14 as the sophisticated spectral fitting using enhanced stellar population models in DR17 results in more accurate measurements of emission lines and kinematic properties, and more accurate ages and metallicities. The DR17 data-analysis pipeline (DAP) has been updated to include better calibrated kinematic maps and tailored analysis options, which allow for finer analysis of kinematic properties \citep[see][for more details]{2021AJ....161...52L}. Most importantly, $\sim 10,000$ galaxies from DR17 include 89 PSBs as compared to the 16 PSBs identified in the $\sim 2800$ galaxies present in DR14, allowing us to perform a more robust comparison between the various samples. We explore the implication of choosing NSA photometry and DR17 instead of DR14 in Appendix \ref{sec: photometry_comparison}.

We obtain the light-weighted ({\tt Age\_LW\_Re\_fit}) and mass-weighted ({\tt Age\_MW\_Re\_fit}) ages, and star formation rates ({\tt log\_SFR\_ssp} and {\tt log\_SFR\_Ha}) for the MaNGA galaxies from the \citet{2018MNRAS.477.4711G} sample cross-matched with the SDSS DR17 pyPipe3D pipeline ({\tt v3\_1\_1}), which is a result of the analysis of stellar populations and ionized gas as presented in \citet{2016RMxAA..52...21S, 2016RMxAA..52..171S, 2018RMxAA..54..217S}. {\tt Age\_LW\_Re\_fit} and {\tt Age\_MW\_Re\_fit} are the logarithms of the LW and MW ages of the stellar population in years within the effective radius of the galaxy, respectively. {\tt log\_SFR\_ssp} is the logarithm of the integrated star formation rate obtained from the stellar mass that formed in the last 32 Myr. {\tt log\_SFR\_Ha} is the logarithm of the integrated star formation rate determined from the integrated H$\alpha$ flux \citep{2016RMxAA..52..171S}. We obtain the stellar mass from the “galSpec” galaxy properties from MPA-JHU \citep{2003MNRAS.341...33K, 2004MNRAS.351.1151B, 2004ApJ...613..898T}. In order to distinguish the MaNGA SFGs from the ETGs, we calculate the specific star formation rate (sSFR; SFR/M$_{\star}$), where SFR is the integrated star formation rate obtained from the stellar mass {\tt log\_SFR\_ssp}. A sSFR $\approx 10^{-11}$ yr$^{-1}$ approximately divides star-forming and quiescent galaxies \citep{2004MNRAS.351.1151B}. This sSFR cut is similar to the H$\alpha$ equivalent-width cut used by \citet{2018ApJ...862....2F} in selecting PSBs. Using this sSFR criteria, we identify 1592 MaNGA SFGs and 916 MaNGA ETGs in the \citet{2018MNRAS.477.4711G} catalog.

We use data products from the MaNGA data-analysis pipeline \citep{2019AJ....158..231W, 2019AJ....158..160B} to measure the angular momentum for the MaNGA SFGs, PSBs and ETGs. We use Marvin tools \citep{2019AJ....158...74C} to obtain the stellar velocity ({\tt stellar\_vel}), stellar velocity dispersion ({\tt stellar\_sigma}), and flux ({\tt spx\_mflux}) for each pixel given as defined by x-coordinate ({\tt spx\_skycoo: on\_sky\_x}) and y-coordinate ({\tt spx\_skycoo: on\_sky\_y}). These velocities have been offset to remove their cosmological redshift with respect to the solar barycentric rest frame. To perform this correction, stellar velocities returned by {\tt pPXF} are used to calculate the observed redshift for individual spaxels, and the effect of the galaxy's input cosmological redshift is removed to convert the velocities to the reference frame of the galaxy \citep{2019AJ....158..231W}. Similarly, the stellar velocity dispersion ($\sigma_{obs}$) is obtained by performing the {\tt pPXF} fit for the template spectra to the MaNGA data. The $\sigma_{obs}$ needs to be corrected for the effective difference in the instrumental dispersion of the template spectra and MaNGA spectra \citep{2019AJ....158..231W}. This correction has been described in Equation (\ref{dispersion correction}).

Example data products for the PSBs are shown in Figures \ref{fig: slow_rotators} and \ref{fig: fast_rotators}, which depict the observed galaxy images from SDSS (leftmost panels), the stellar velocity (middle panels), and the stellar velocity dispersion (rightmost panels). Figure \ref{fig: slow_rotators} shows four slow rotators and Figure \ref{fig: fast_rotators} shows four fast rotators from the PSB sample that were distinguished using the classification discussed in Section \ref{sec: results}. These figures showcase the diversity in the PSB sample, as galaxies like 11752-3701 (slow rotator) and 9493-1902 (fast rotator), the top galaxies in each figure, have similar morphologies. However, they vary quite a bit in their kinematics, as depicted by the corresponding stellar velocity and dispersion plots. On the other hand, 12071-3702 (slow rotator) and 8080-3704 (fast rotator) are highly elliptical galaxies, and galaxies like 7960-12704 (fast rotator) are disk-like and have spiral arms. The PSB sample also displays a range of merger signatures and tidal features, like 11753-3704 (fast rotator). These characteristics are quite evident in not just the SDSS images in the left panels, but also from the inner kinematics presented via the stellar velocity and stellar velocity dispersion in the middle and right panels.

\begin{table*}[!ht]
    \centering
    \caption{MaNGA galaxy properties\footnote{This table is available in its entirety in machine-readable form in the \href{https://doi.org/10.3847/1538-4357/ada7f6}{online article}.}}
    \begin{tabular}{lll}
    \hline 
        Name & Units & Description \\ \hline
        Plateifu & ... & Marvin Plate ID to obtain spectroscopic data \citep{2022ApJS..259...35A} \\ 
        MaNGA\_Id & ... &  MaNGA ID for the galaxy \citep{2022ApJS..259...35A} \\
        RA & deg & R.A. of the center of the galaxy \citep{2011ApJS..193...29A}  \\
        Dec & deg & Decl. of the center of the galaxy \citep{2011ApJS..193...29A} \\
        Epsilon & ... & Ellipticity ($\epsilon = $1 - {\tt SERSIC\_BA}) of the disk \citep{2011AJ....142...31B} \\
        FWHM\_PSF & arcsec & Full-width at half-maximum (dispersion)\footnote{Given by the point-spread function, which is assumed to be a Gaussian} (zfwhm) \\
        Sérsic\_n & ... & Sérsic index ({\tt SERSIC\_N}) \citep{2011AJ....142...31B} \\
        R\_eff & arcsec & Half-light radius ({\tt SERSIC\_TH50}) of the galaxy \citep{2011AJ....142...31B} \\
        Semi\_Maj\_axis & kpc & Semi-major axis (= {\tt SERSIC\_TH50}/($1-\sqrt{\epsilon}$) of the galaxy \\
        Phot\_angle & deg & Position angle ({\tt SERSIC\_PHI}) of the disk east of north \citep{2011AJ....142...31B} \\
        Lambda\_Re & ... & Corrected angular momentum at half-light radius \\
        Stellar\_mass & M$_\odot$ & Stellar mass from galSpec MPA-JHU \citep{2003MNRAS.341...33K} \\
        SFR\_ssp & M$_\odot$ yr$^{-1}$ & Star formation rate ({\tt log\_SFR\_ssp}) of the galaxy \citep{2016RMxAA..52...21S, 2016RMxAA..52..171S, 2018RMxAA..54..217S} \\
        SFR\_ha & M$_\odot$ yr$^{-1}$ & Star formation rate ({\tt log\_SFR\_Ha}) of the galaxy in H$\alpha$ \citep{2016RMxAA..52...21S, 2016RMxAA..52..171S, 2018RMxAA..54..217S} \\
        LW\_age & Gyr & Luminosity-weighted age ({\tt Age\_LW\_Re\_fit}) of the galaxy \citep{2016RMxAA..52...21S, 2016RMxAA..52..171S, 2018RMxAA..54..217S} \\
        MW\_age & Gyr & Mass-weighted age ({\tt Age\_MW\_Re\_fit}) of the galaxy \citep{2016RMxAA..52...21S, 2016RMxAA..52..171S, 2018RMxAA..54..217S} \\
        Asymmetry & ... & Morphological asymmetry ({\tt A}) of the galaxy \\
        ~ & ~ & \citep{2010AJ....139.2525H, 2011MNRAS.412..727C, 2022MNRAS.512.2222V} \\
        Type & ... &  Type of galaxy (SFG, PSB or ETG)\\
    \hline
    \end{tabular}
    \label{Tab: manga_data}
\end{table*}

\subsection{Other Sources of Data}
\label{sec: other_data}

We obtain the data for a different sample of 260 ETGs from the ATLAS$^{3D}$ Project \citep{2011MNRAS.413..813C}. This project obtained its data from the SAURON IFU on the William Herschel Telescope (WHT), combined with deep imaging done with MegaCam on the Canada-France-Hawaii Telescope.

We use data sets provided by multiple papers under the umbrella of the ATLAS$^{3D}$ project to obtain parameters that enable us to make comparisons for angular momentum in the context of LW age, MW age, and stellar mass. We acquire ellipticity ({\tt Epsilon\_e}) and angular momentum ({\tt LambdaR\_e}) at effective radius from \citet{2011MNRAS.414..888E}. \citet{2011MNRAS.414..888E} calculated the $\lambda_{Re}$ using the expression from \citet{2007MNRAS.379..401E}, which is given in Equation (\ref{lambda_eq}). 

We obtain the stellar masses for each galaxy using the luminosity ({\tt logLum}) from \citet{2013MNRAS.432.1709C} and the mass-luminosity ratios ({\tt logML\_star}) from \citet{2013MNRAS.432.1862C} in the SDSS $r$ band. {\tt logLum} is the analytic total luminosity of the multi-Gaussian expansion model for an assumed distance and extinction (given in \citet{2011MNRAS.413..813C}). {\tt logML\_star} is the mass-to-light ratio for the stellar component within $R_e$ for the same assumed distance and extinction for the best-fitting Jeans anisotropic model with a Navarro-Frenk-White halo profile from \citet{2013MNRAS.432.1709C, 2013MNRAS.432.1862C}. We procure the LW age ({\tt Age\_SSP}) and MW age ({\tt Age\_SHF}) within $R_e$ from \citet{2015MNRAS.448.3484M}. The LW age is estimated using the single stellar population (SSP) models from \citet{2007ApJS..171..146S} and the measured age-sensitive Lick H$\beta$ index. The MW age is computed using the weight ($w_i$) given by the {\tt pPXF} fit to each $i$th template which has age $t_{SSP,i}$ \citep{2015MNRAS.448.3484M}.

\begin{table*}
    \centering
    \caption{ETG properties for the ATLAS$^{3D}$ Comparison Sample\footnote{This table is available in its entirety in machine-readable form in the \href{https://doi.org/10.3847/1538-4357/ada7f6}{online article}.}}
    \begin{tabular}{lll}
    \hline 
        Name & Units & Description \\ \hline
        Galaxy & ... & Name of the galaxy \citep{2011MNRAS.414..888E} \\
        Lambda\_obs & ... & Specific angular momentum as calculated by \citet{2011MNRAS.414..888E} \\
        Epsilon & ... & Ellipticity ({\tt Epsilon\_e}) from \citep{2011MNRAS.414..888E} \\
        FWHM\_PSF & arcsec & Full-width at half-maximum (dispersion)\footnote{Used an average value of 1.25$\arcsec$ \citep{2013MNRAS.432.1894S}} \\
        Sérsic\_n & ... & Sérsic index ({\tt n}) of the galactic disk \citep{2013MNRAS.432.1768K} \\
        Semi\_Maj\_axis & kpc & Semimajor axis calculated using {\tt logRe\^maj} and {\tt D} of the galaxy \citep{2013MNRAS.432.1709C} \\
        Lambda\_corr & ... & {\tt Lambda\_obs} corrected for beam-smearing \citep{2018MNRAS.477.4711G} \\
        Stellar\_mass & M$_\odot$ & Stellar mass calculated using {\tt logLum} \citep{2013MNRAS.432.1709C} and \\
        ~ & ~ & {\tt logML\_star} \citep{2013MNRAS.432.1862C} as described in Section \ref{sec: other_data}\\
        LW\_age & Gyr & Luminosity-weighted age ({\tt Age\_SSP}) of the galaxy \citep{2015MNRAS.448.3484M} \\
        MW\_age & Gyr & Mass-weighted age ({\tt Age\_SHF}) of the galaxy \citep{2015MNRAS.448.3484M} \\
        \hline
    \end{tabular}
    \label{Tab: etg_data}
\end{table*}


\section{Results}

\subsection{Angular Momentum Calculation and Correction}
\label{sec: calculation}

\begin{figure*}
\plotone{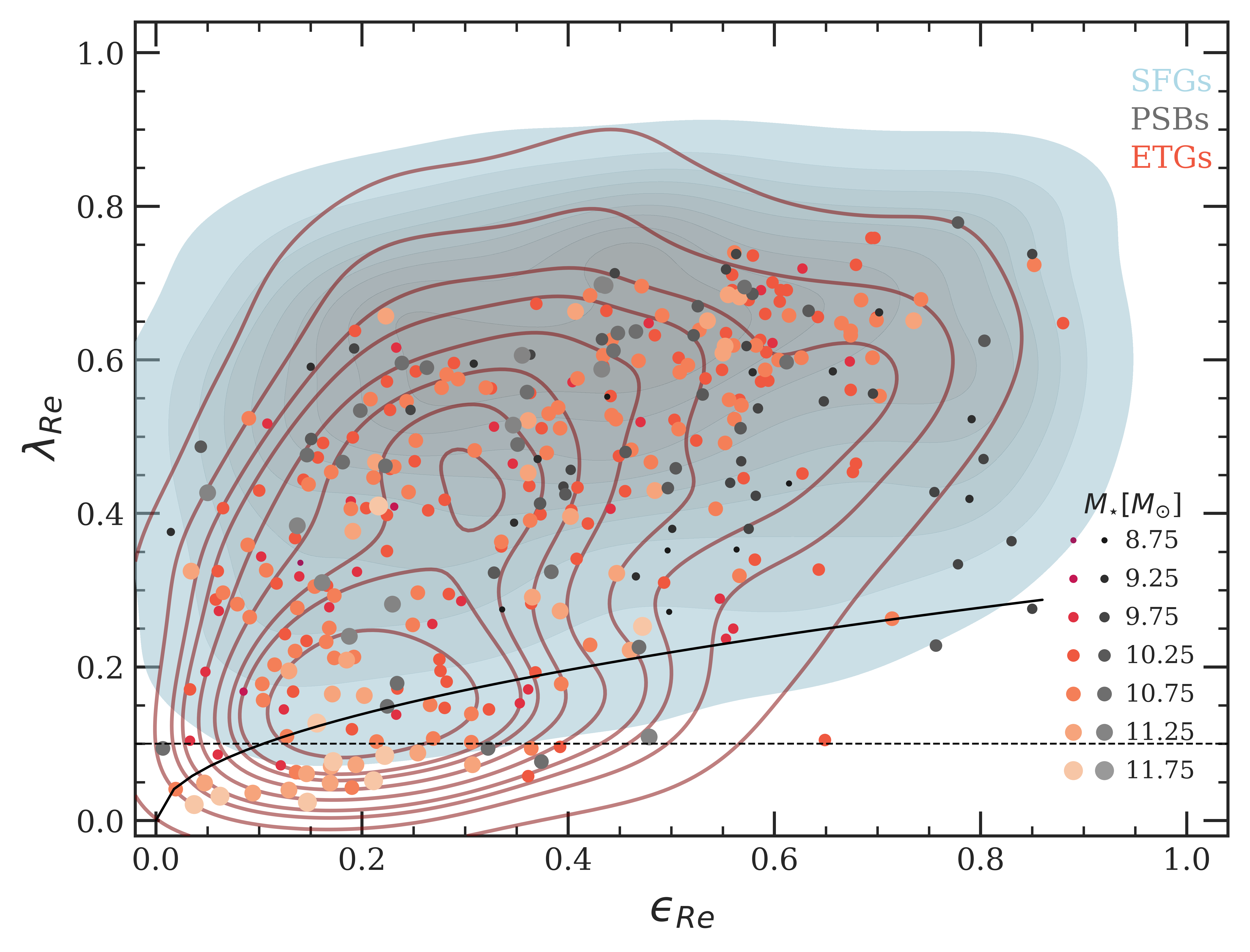}
\caption{Angular momentum at half-light radius, $\lambda_{Re}$ vs. ellipticity at half-light radius, $\epsilon_{Re}$, for 89 PSBs from the MaNGA Survey and 260 ETGs from the ATLAS$^{3D}$ Project, with point sizes scaled by log stellar mass. The blue and red contour lines represent the 1592 SFGs and 916 ETGs from MaNGA, respectively. The black dashed line at $\lambda_{Re}=0.1$ separates the slow and fast rotators according to the classical definition, while the black solid line given by $0.31 \times\sqrt{\epsilon_{Re}}$ \citep{2011MNRAS.414..888E} performs a similar function as the black dashed line, but takes into account the ellipticity of the galaxies in the classification of slow and fast rotators. We find that a higher fraction of ETGs are slow rotators than the PSB and SFG samples, with the SFG sample having significantly higher average $\lambda_{Re}$.
\label{fig: lambda_vs_ellipticity}}
\end{figure*}

We calculate the specific angular momentum for each galaxy using the expression from \citet{2007MNRAS.379..401E} as
    \begin{equation}\label{lambda_eq}
        \lambda_R = \frac{\sum_{i=1}^{N_p} F_i R_i |V_i|}{ \sum_{i=1}^{N_p} F_i R_i\sqrt{{V_i}^2 + {\sigma_i}^2}}
    \end{equation}
where $F_i$, $R_i$, $V_i$, and $\sigma_i$ are the flux, distance from the center, stellar velocity, and dispersion respectively in the $i$th pixel. For the MaNGA data, we utilize the Python functions provided by \citet{2018MNRAS.477.4711G}.

We calculate the angular momentum (given in column (3) of Table \ref{Tab: manga_data}) of the 89 PSBs, 1592 SFGs and 916 ETGs using Equation (\ref{lambda_eq}) and the {\tt lambdaR\_e\_calc} function defined by \citet{2018MNRAS.477.4711G} by passing the stellar velocity, stellar velocity dispersion, and flux in each pixel (Section \ref{sec: data_and_methods}). We also provide effective radius, ellipticity, position angle, full width at half-maximum (FWHM), and Sérsic index (all of which are described in Table \ref{Tab: manga_data}) from the NSA catalog, in order to determine the 1 $R_e$ ellipse in which to calculate $\lambda$. The stellar velocity dispersion has been corrected for the instrumental dispersion ({\tt stellar\_sigmacorr}) using the given correction provided by \citet{2019AJ....158..231W}:
\begin{equation}\label{dispersion correction}
    \sigma_{corrected} = \sqrt{\sigma_{raw}^2 - \sigma_{corr}^2}
\end{equation}
According to \citet{2018MNRAS.477.4711G}, the finite point-spread function (PSF) results in the smearing of the velocity field. To correct for this effect, they provide an optional analytic correction as part of their {\tt lambdaR\_e\_calc} function. This correction can be applied to regular rotators whose semimajor axis is larger than the PSF dispersion and Sérsic index is in the range $0.5-6.5$. Thus, we obtain the beam-corrected angular momentum for the MaNGA galaxies. All the parameters used and the angular momentum calculated for the MaNGA SFGs, PSBs and ETGs are presented in Table \ref{Tab: manga_data}.

We use the {\tt lambdaR\_e\_correct} function provided by \citet{2018MNRAS.477.4711G} to correct the specific angular momentum of the ETG sample (from \citet{2011MNRAS.414..888E}) for beam-smearing. To do so, we provide the uncorrected angular momentum, semimajor axis ({\tt log R$^{maj}_e$} from \citet{2013MNRAS.432.1709C}), PSF dispersion ({\tt sig\_PSF}) and Sérsic indices ({\tt n} from \citet{2013MNRAS.432.1768K}). We ensure that the Sérsic indices are within the range $0.5-6.5$ by replacing them with 0.5001 if $n<0.5$ and with 6.4999 if $n>6.5$. This replacement for Sérsic indices was necessary for the ATLAS$^{3D}$ galaxies as 20 of the 260 ETGs had Sérsic indices outside the range of $0.5-6.5$. The dispersion in the PSF is estimated to be $\sigma_{PSF} \approx 1.25\arcsec/2.355$ \citep{2018MNRAS.477.4711G}, where 1.25$\arcsec$ is the average FWHM for the ETG sample given by \citet{2013MNRAS.432.1894S}. Since the typical seeing for the ETG sample (FWHM $\approx 1.25\arcsec$) is better than the seeing of MaNGA (FWHM $\approx 2\arcsec$), the beam-smearing is lesser in comparison for the former, thus, averaging the FWHM should not change the results significantly. The negligible effect of the averaged FWHM on the ETG sample is evident when our results for the percentage of slow and fast rotators in the sample do not change even if we do not correct the angular momentum for beam-smearing. All the aforementioned parameters, among others, for the ATLAS$^{3D}$ ETG sample are presented in Table \ref{Tab: etg_data}.

\subsection{Identifying Slow and Fast Rotators}
\label{sec: results}

We plot the beam-corrected specific angular momentum at half-light radius for 89 MaNGA PSBs and 260 ETGs from ATLAS$^{3D}$ against the ellipticity at half-light radius in Figure \ref{fig: lambda_vs_ellipticity}. The points are scaled in size and color by stellar mass (acquired as described in Section \ref{sec: sample_selection} and  Section \ref{sec: other_data} for PSBs and ETGs, respectively), with the corresponding mass given in the legend. The blue and red contours depict the 1592 SFG and 916 ETG MaNGA galaxies, respectively. We find that a higher fraction of ETGs are slow rotators than the PSB and SFG samples, with the SFG samples having significantly higher average $\lambda_{Re}$.

As shown in Figure \ref{fig: lambda_vs_ellipticity}, we show two criteria to separate the slow and fast rotators: the classical $\lambda_{Re} = 0.1$ (black dashed line) boundary condition and the curve given by $0.31 \times\sqrt{\epsilon_{Re}}$ (black dotted line) \citep{2011MNRAS.414..888E}. \citet{2011MNRAS.414..888E} found that the specific angular momentum generally increases with the ellipticity, even for non-regular rotators, so they developed a relationship between $\lambda_{Re} - \epsilon_{Re}$ to distinguish slow and fast rotators. This relationship also accounts for the increasing galactic anisotropy with ellipticity. Thus, $\lambda_{Re} = 0.31 \times\sqrt{\epsilon_{Re}}$ is a better measure of whether a galaxy is a slow or fast rotator. Using this as the preferred limiting condition, we find that 5 out of 89 ($\sim$6\%) PSBs, 36 out of 260 ($\sim$14\%) ETGs from ATLAS$^{3D}$, 56 out of 1592 ($\sim$3.5\%) MaNGA SFGs, and 196 out of 916 ($\sim$19.5\%) MaNGA ETGs are slow rotators. We also find that $\sim42\%$ of MaNGA SFGs, $\sim25\%$ of MaNGA PSBs, $\sim18\%$ of MaNGA ETGs and $\sim21\%$ of ATLAS$^{3D}$ ETGs have higher angular momentum than $\lambda_{Re}=0.6$. Furthermore, within the MaNGA sample, $\sim 64\%$ SFGs have higher momentum than the median $\lambda_{Re} \approx 0.5$ of the fast-rotator PSBs, and $\sim 74\%$ SFGs and $\sim 62\%$ PSBs rotate faster than the median $\lambda_{Re} \approx 0.44$ of the fast-rotator ETGs. The median $\lambda_{Re} \approx 0.57$ of the SFGs is higher than the median $\lambda_{Re}$ of the fast-rotator PSBs and ETGs, respectively. Thus, the ETGs generally have lower angular momentum as compared to the PSBs, which tend to rotate slower than the SFGs. 

\begin{figure}
\includegraphics[width=0.49\textwidth]{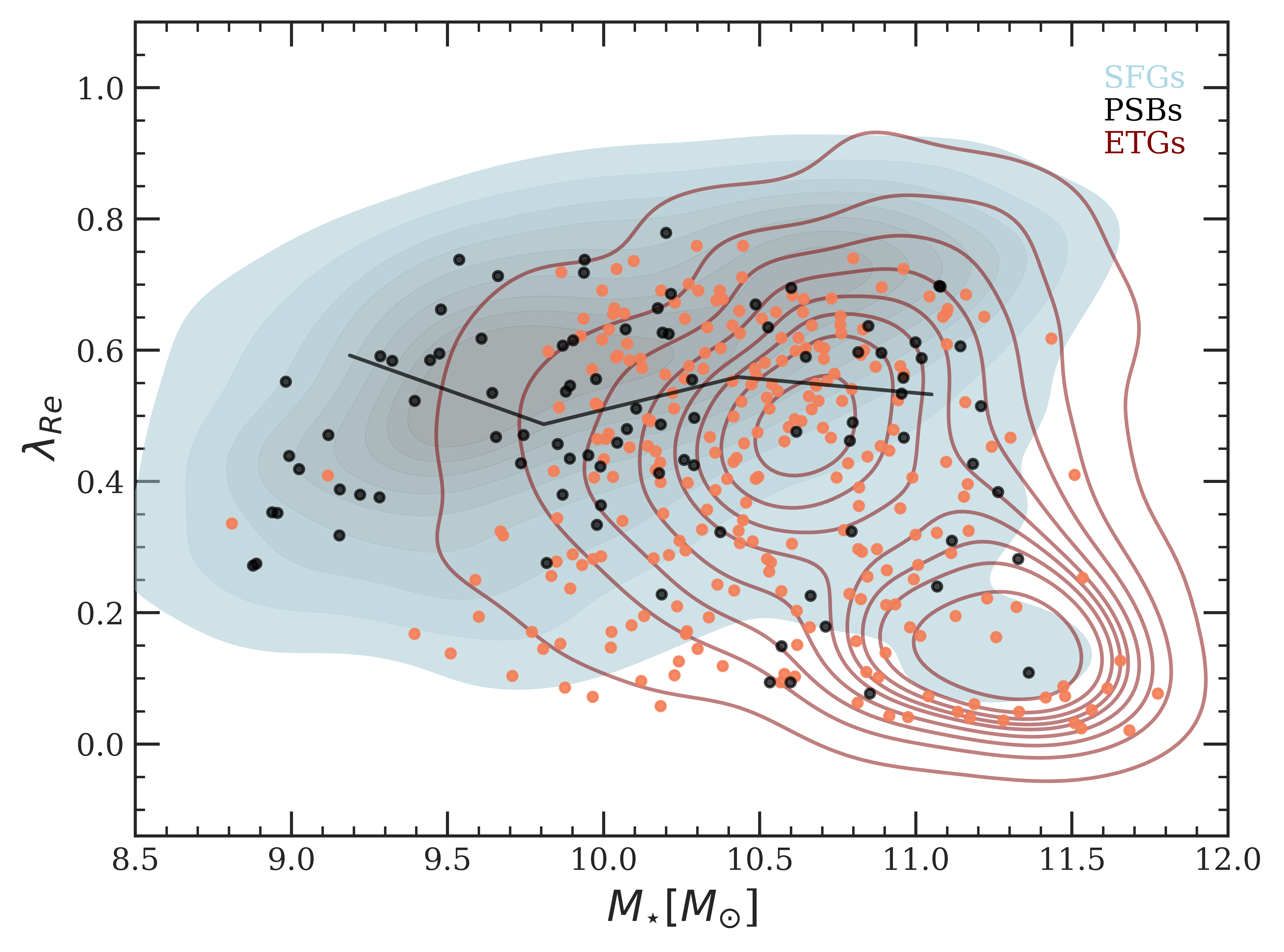}
\caption{Angular momentum vs. stellar mass for 89 PSBs (black scatter), 1592 SFGs (blue contours) and 916 ETGs (red contours) from MaNGA, and 260 ATLAS$^{3D}$ ETGs (red scatter). The black line represents the running average for the PSBs. We find that galaxies, especially the early types, with higher stellar mass tend to rotate slowly as compared to galaxies with lower stellar mass. While most of the slow-rotator ETGs are among the most massive ETGs, a similar effect is not seen for the PSB and SFG samples. We explore this effect quantitatively in the text. This implies that the high-stellar-mass slow-rotator ETGs have experienced further growth and loss of angular momentum since they may have passed through the PSB phase.
}
\label{fig: lambda_vs_mass}
\end{figure}

The scaling and coloring of points based on stellar mass shows that galaxies, especially ETGs, with higher stellar mass prefer to rotate slower as compared to their counterparts. We explore trends in angular momentum with stellar mass in Figure \ref{fig: lambda_vs_mass}. In this figure, we can see a distinct population of high-stellar-mass slow rotators in the ETG sample that is not present for the PSBs and SFGs. To quantify this observation, we consider the fraction of slow rotators in the top 30th percentile of stellar mass for each galaxy type: $\sim32\%$  of MaNGA SFGs, $40\%$ of MaNGA PSBs, $\sim58\%$ of MaNGA ETGs and $\sim58\%$ of ATLAS$^{3D}$ ETGs. While the slow-rotator SFG and PSB galaxies are slightly overrepresented in the highest-stellar-mass galaxies, the slow-rotator ETGs are significantly more biased towards the higher-mass end. We find similar levels of overrepresentation when we consider the top 10th or 20th percentiles of stellar mass.

We consider the fraction of slow rotators in the PSBs within the different parent samples. Of the 89 PSBs, there are 24 E+As, 19 SPOGs, and 74 PCAs (with significant overlap). Two of the five slow rotators are in multiple classifications, and we find one E+A, one SPOG, and five PCA slow rotators. Within the small number uncertainties, the fraction of slow rotators is consistent between each parent PSB sample. 

\section{Discussion}
\label{sec: discussion}

\subsection{Evolution of Angular Momentum with Age}
\label{sec: evolution_with_age}

\begin{figure*}
\begin{center}
\includegraphics[width=0.49\textwidth]{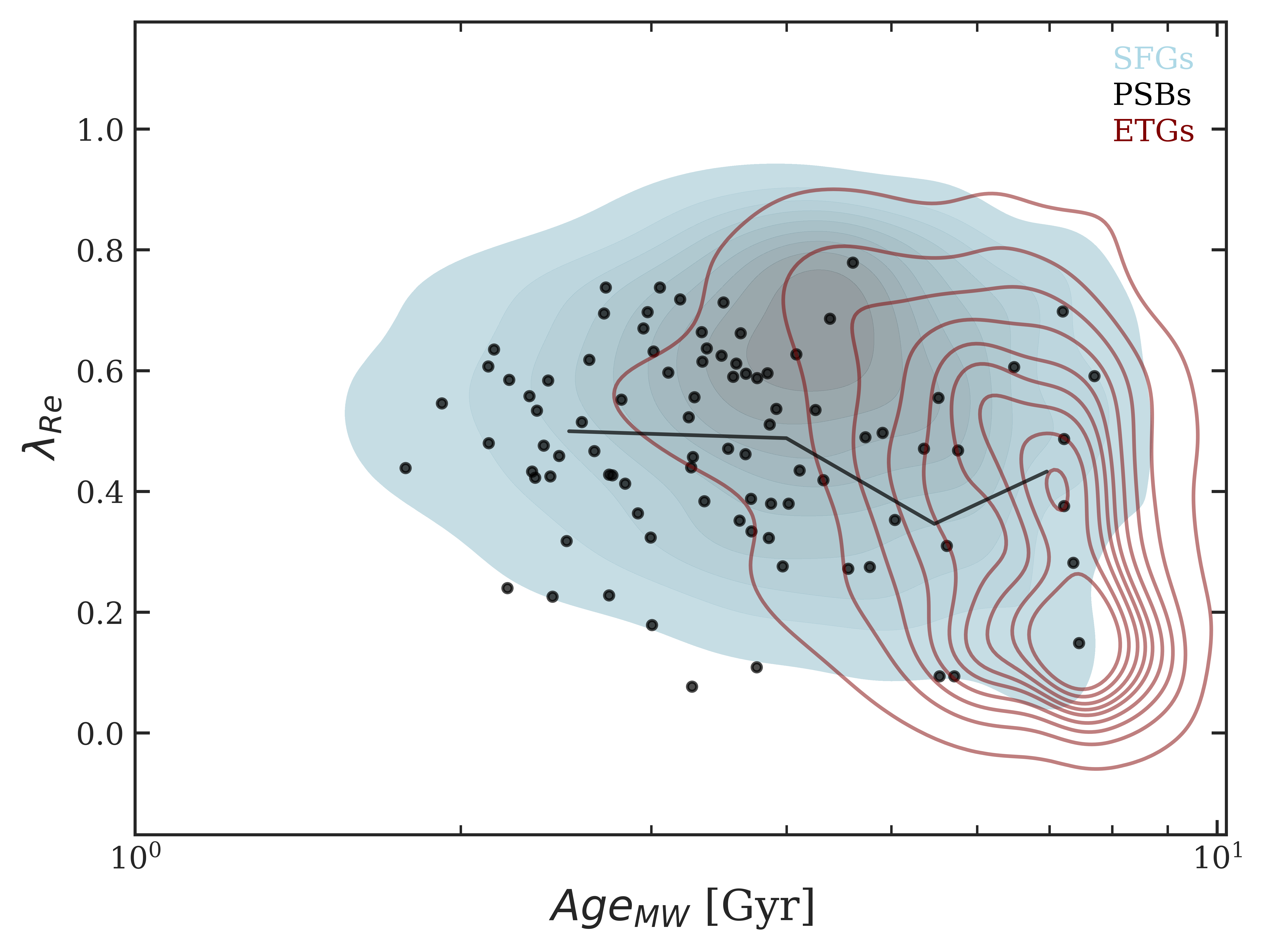}
\includegraphics[width=0.49\textwidth]{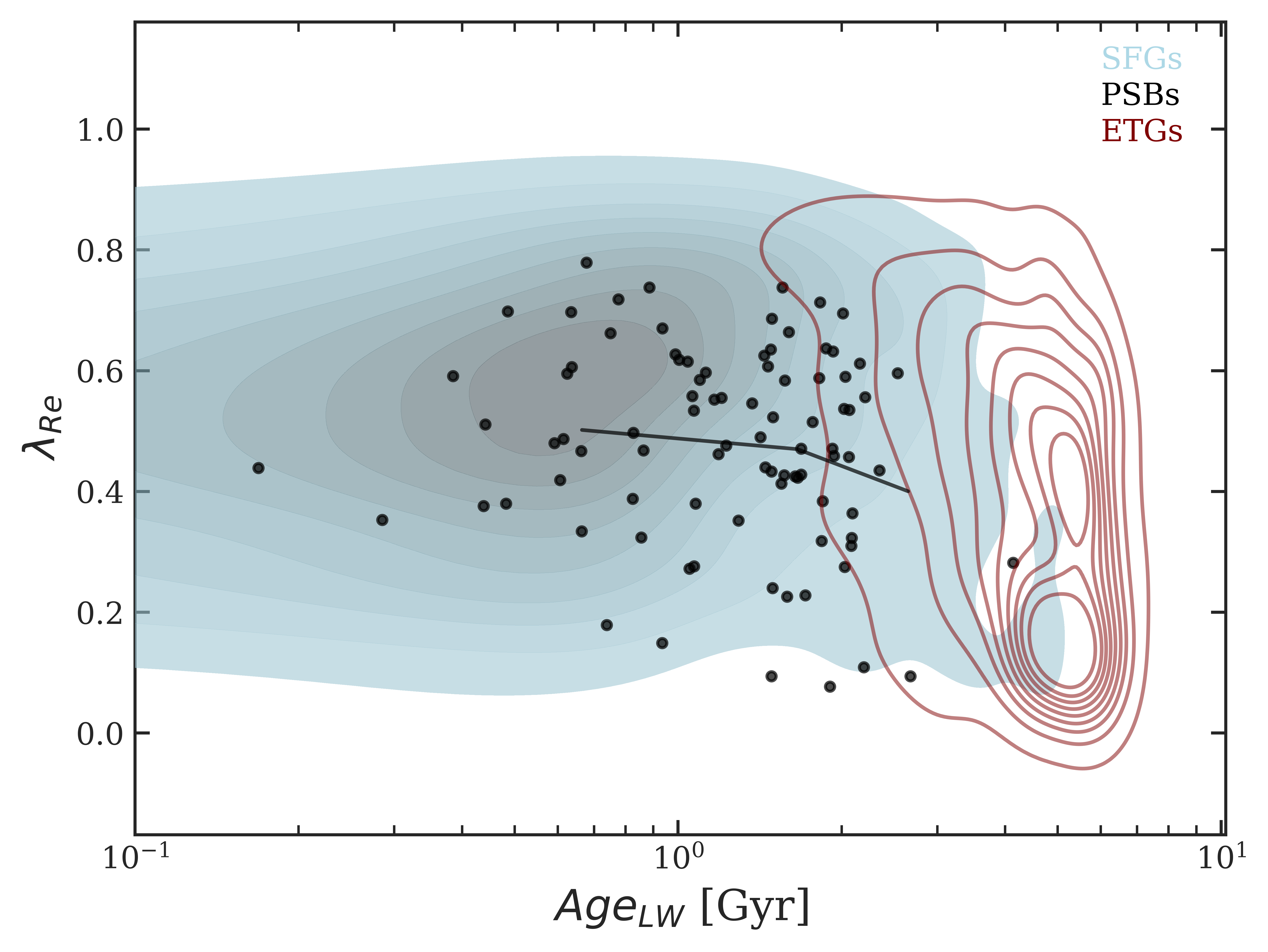}
\end{center}
\caption{Left: angular momentum at half-light radius, $\lambda_{Re}$, vs. mass-weighted (MW) age for 89 PSBs (black points), 1586 SFGs (blue contours), and 914 ETGs (red contours) from the MaNGA survey. The black line represents the running average for PSBs. We observe a clear distinction in age between the SFG and ETG samples, with the median MW age of the PSB sample ($\sim$3.4 Gyr) younger than the median MW ages for the SFG ($\sim$4.6 Gyr) and ETG ($\sim$6.9 Gyr) samples. A significant anticorrelation between MW age and $\lambda_{Re}$ is seen for the SFG and ETG samples, though not within the PSB sample. At MW ages $\sim$5 Gyr, where we observe an overlap between the older PSBs and the younger ETGs, the PSB and ETG distributions in angular momentum are quite similar. For older ETGs, we see that a trend towards lower angular momentum continues. This implies that even after the PSBs evolve into ETGs, they do not stop shedding off their angular momentum. Right: angular momentum at half-light radius, $\lambda_{Re}$, vs. luminosity-weighted (LW) age for 89 PSBs (black points), 1586 SFGs (blue contours), and 914 ETGs (red contours) from the MaNGA survey. The black line represents the running average for the PSBs. Similar to Figure \ref{fig: lambda_vs_age}(a), we observe a clear distinction between the SFG and ETG samples, though now in the case of LW age, the PSB ages are typically between the SFG and ETG samples. Quantitatively, the median LW age of the PSB sample ($\sim$1.45 Gyr) is older than the median LW age for the SFGs ($\sim$0.9 Gyr) and younger than the median LW age of the ETGs ($\sim$4.9 Gyr). For the LW ages, we observe a significant trend within the ETG samples, but no significant trend in the SFG and the PSB sample. The fact that the angular momentum is more closely coupled with the MW and LW age for the SFGs and ETGs as compared to the PSBs implies that the kinematic evolution is more closely coupled to the longer-scale formation history and evolution of the galaxy in its star-forming and early-type phases, instead of the time it spends in the post-starburst phase.}
\label{fig: lambda_vs_age}
\end{figure*}

To further our understanding of the evolution of angular momentum across the three classes, we study its evolution with the age of the stellar populations. To do so, we plot the angular momentum against MW and LW ages (as shown in Figure \ref{fig: lambda_vs_age}). As discussed in Section \ref{sec: MaNGA}, we obtain the respective ages for MaNGA SFGs, PSBs and ETGs from the SDSS DR17 pyPipe3D pipeline ({\tt v3\_1\_1}), which is a result of the analysis of stellar populations and ionized gas as presented in \citet{2016RMxAA..52...21S, 2016RMxAA..52..171S, 2018RMxAA..54..217S}. 

We observe that angular momentum decreases with increasing age across all three samples for MW and LW age, with the early types having older MW and LW ages and more slow rotators than PSBs and SFGs. While the PSBs are on average younger than both the SFGs and ETGs based on their median MW age, their median LW age seems to be older than the average SFG. The younger MW age of the PSBs as compared to the SFGs and ETGs can be attributed to the significant recent burst populations in these galaxies, while the recent quenching in the PSBs leads to their older LW ages as compared to the SFGs. At MW ages $\sim$5 Gyr, where the older PSBs have similar MW ages to the younger ETGs, the PSB and ETG distributions in angular momentum overlap. For older ETGs, we see that a trend toward lower angular momentum continues. This implies that even after the PSBs evolve into ETGs, they do not stop shedding off their angular momentum. To quantify this, we compute the Spearman correlation (presented in Table \ref{tab: spearman}) for all three classes of galaxies for MW and LW age. We find significant anticorrelations of angular momentum with MW age for the SFG and ETG samples, though not within the PSB sample. For the LW age, we do not observe a significant anticorrelation for the PSB sample as well. The fact that the angular momentum is more closely coupled with the MW and LW ages for the SFGs and ETGs as compared to the PSBs implies that the kinematic evolution is more closely coupled to the longer-scale formation history and evolution of the galaxy in its star-forming and early-type phases, instead of the time it spends in the post-starburst phase.

We also test whether we see evolution within the PSB phase by testing the correlation between angular momentum and age since the starburst ended using the stellar population fits by \citet{2018ApJ...862....2F}. Consistent with the lack of correlation in MW or LW ages, we observe no significant trend within the PSBs in angular momentum versus starburst age.

\begin{table}[!ht]
\centering
\caption{Spearman Result for MW and LW Ages for MaNGA SFGs, PSBs, and ETGs}
\label{tab: spearman}
\begin{tabular}{lll}
\hline
Galaxy & MW age & LW age \\ \hline
SFGs & ($\rho \approx$ -0.12, p $\approx$ $10^{-6}$) & ($\rho \approx$ $0.11$, p $\approx 10^{-5}$) \\
PSBs & ($\rho \approx$ -0.13, p $\approx$ 0.22) & ($\rho \approx$ -0.15, p $\approx$ 0.16) \\
ETGs & ($\rho \approx$ -0.30, p $\approx 10^{-20}$) & ($\rho \approx$ -0.28, p $\approx 10^{-18}$)\\
\hline
\end{tabular}
\end{table}

\subsection{Evolution of Angular Momentum with Asymmetry}
\label{sec: evolution_with_asymmetry}

\begin{figure}
\includegraphics[width=0.49\textwidth]{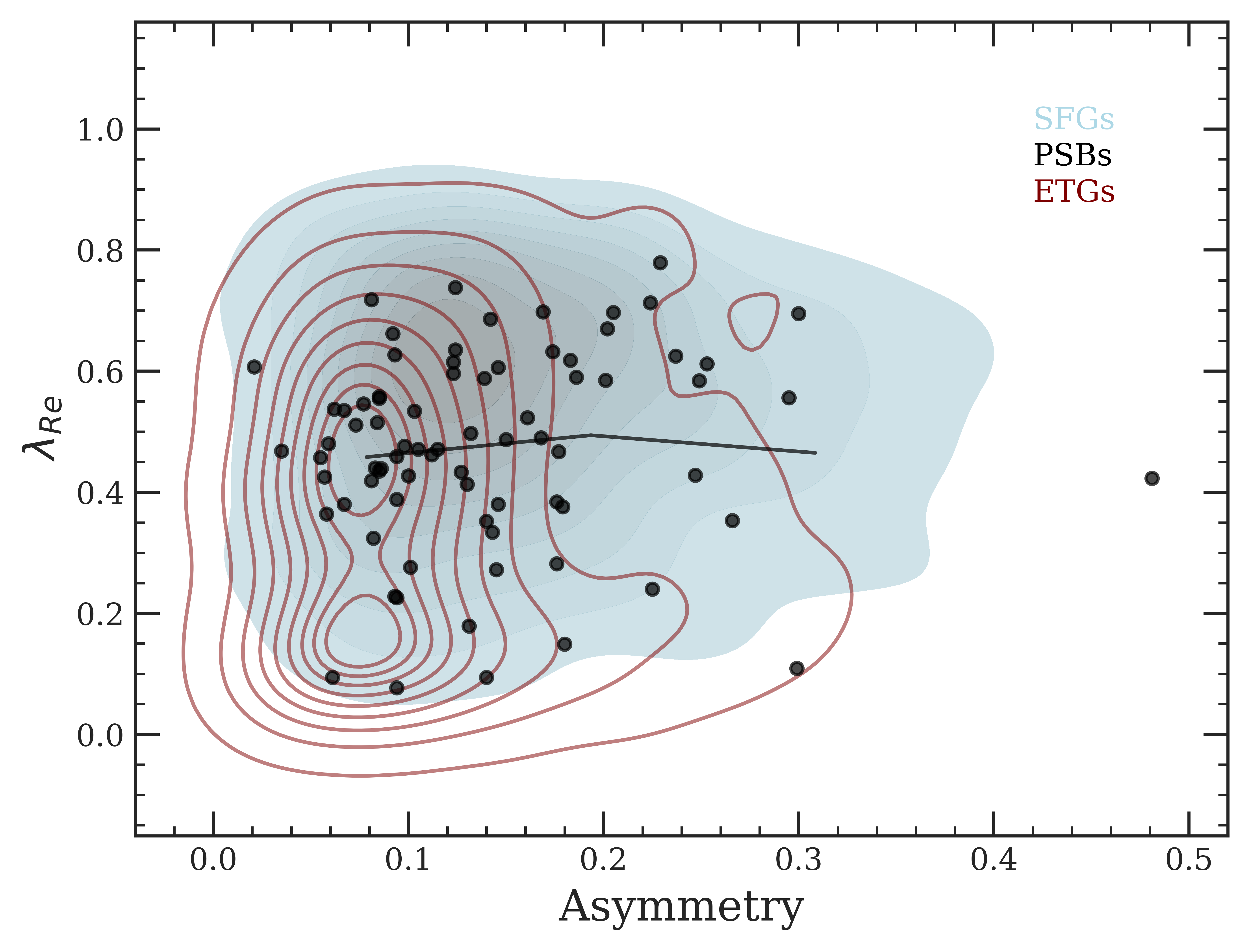}
\caption{Angular momentum at half-light radius, $\lambda_{Re}$, vs. asymmetry, $A$, for 79 PSBs (black points), 1341 SFGs (blue contours), and 841 ETGs (red contours) from the MaNGA survey. The black line represents the running average for the PSBs. The ETGs tend to be fast and slow rotators for similar asymmetry values, while angular momentum is correlated with asymmetry for the SFGs. The PSBs lie between the ETGs and SFGs in both angular momentum and asymmetry. We do not observe a significant trend between angular momentum and asymmetry within the PSB sample. Thus, if PSBs are more likely to be slow rotators after a major merger, the rapid decay of observable tidal features may have obscured this effect.
}
\label{fig: lambda_vs_asy}
\end{figure}

Asymmetric features in galaxies can be used to trace their recent merger history due to the presence of tidal tails, shells, and companion galaxy remnants, as well as to separate galaxies between morphological types. The rotational asymmetry $A$, a purely morphological parameter, is calculated by rotating a galaxy by 180$^{\circ}$ and subtracting it from the original image, and further applying relevant corrections to take into account background noise, etc.

PSBs tend to exhibit high asymmetry compared to the ETGs, which could be due to tidal/relic features associated with the recent starburst and/or recent interaction with another galaxy or dust lanes in the case of dusty PSBs \citep{2008ApJ...688..945Y, 2021ApJ...919..134S}. The SFGs have high $A$ due to the presence of spiral arms and flocculence in the disk. However, the ETGs tend to have low $A$ due to the absence of tidal features or spiral arms.

The asymmetry thus provides a way to test for the effects of recent mergers and interaction in galaxies, and test for trends with angular momentum content. We would expect the PSBs resulting from major mergers to be characterized by significantly lower angular momentum, however the expected trends depend on the relative timescales for tidal features to be present and angular momentum to be lost. Low values of $A$ could be present in post-major-merger galaxies due to the fact that their merger features fade quickly as the PSBs age \citep{2016MNRAS.456.3032P}. The PSBs also have a large range of angular momentum, which can obscure any subtle trends we might observe.

Asymmetry values for the PSBs, ETGs, and SFGs in the MaNGA survey are obtained from the MaNGA Visual Morphology Catalogue \citep{2022MNRAS.512.2222V}. Following digital image postprocessing, the catalogue detects the inner structures and external, low-surface-brightness characteristics. The morphological classification process was implemented empirically, based on the methods described in \citet{2010AJ....139.2525H} and \citet{2011MNRAS.412..727C}. They reported an estimation of the CAS parameters (concentration, csymmetry, and clumpiness) derived from the Legacy Surveys and SDSS imaging, following the method by \citet{2003ApJS..147....1C}.

Figure \ref{fig: lambda_vs_asy} compares the angular momentum against rotational asymmetry for 78 PSBs, 1341 SFGs, and 841 ETGs from the MaNGA survey. The black line is the running average for the PSBs. We observe that the slow and fast rotators in the ETG sample are concentrated around small values of asymmetry, however the fast rotators in the SFG sample tend to exhibit high asymmetry values. The finding of higher angular momentum in the SFGs with more asymmetry may be due to a similar effect as seen by \citet{2012ApJS..203...17R}. They find that later Hubble-type galaxies and galaxies with lower bulge-to-total light ratios have more angular momentum, even within the class of spirals. The PSBs, on the other hand, do not seem to exhibit a correlation (Spearman correlation: $\rho \approx$ 0.15, p $\approx$ 0.21) for angular momentum and asymmetry. Thus, if PSBs are more likely to be slow rotators after a major merger, the rapid decay of observable tidal features may have obscured this effect.

Deeper photometry may reveal more subtle trends between the asymmetry caused by merger features and the effect on angular momentum. \citet{2024MNRAS.tmp..386R} used deep imaging from Subaru Hyper Suprime-Cam to test for the presence of tidal features and shells around galaxies as a function of angular momentum, finding a correlation between the presence of shells and lower angular momentum in a subset of the younger galaxies. Deeper photometry in next-generation surveys such as the Rubin Observatory LSST will provide useful constraints on the correlation of merger features with angular momentum during the PSB phase and into the ETG phase.

\subsection{Testing for Selection Bias in the PSB Sample}

\begin{figure}
\begin{center}
\includegraphics[width=0.49\textwidth]{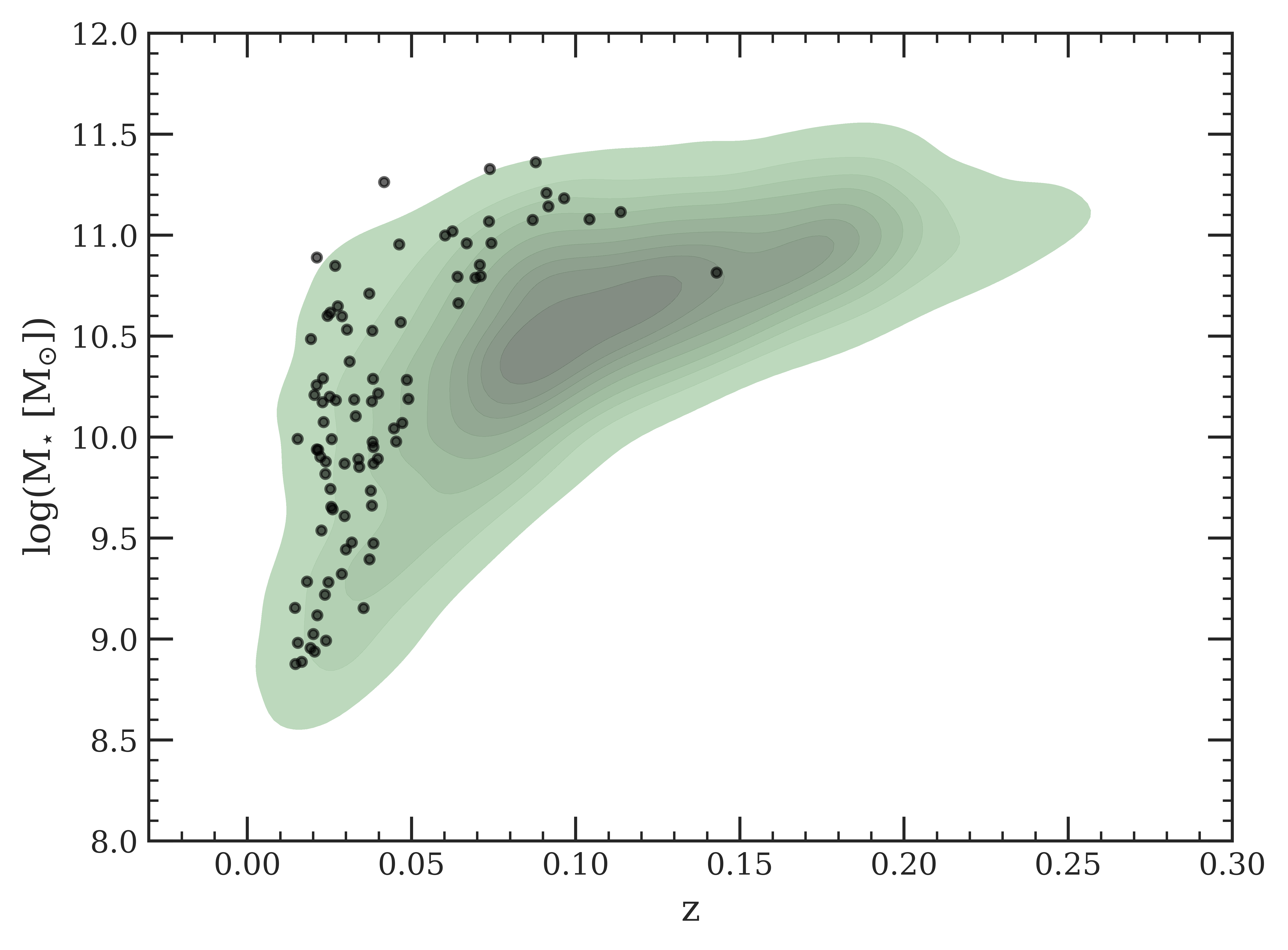}
\end{center}
\caption{We test for dependence between redshift and stellar mass by comparing the smaller PSB sample used in this work with the combined PSB sample from \citet{2023ApJ...950..153F}. The black points are the 89 MaNGA PSBs while the green contours represent the 4985 PSBs from the combined sample. We find that despite the PSBs preferring lower redshifts, they span the entire stellar mass range of the bigger sample.}
\label{fig: mass_vs_z}
\end{figure}

We consider the range in properties of the MaNGA PSBs compared to the parent PSB sample in order to assess biases in this sample that may affect our conclusions. The MaNGA survey targets low-redshift galaxies, so we must ensure this does not result in biases in the trends with angular momentum. In Figure \ref{fig: mass_vs_z}, we compare the redshift and stellar mass of the 89 MaNGA PSBs (black points) to those of the parent sample of $\sim$5000 PSBs (green contours) from \citet{2023ApJ...950..153F}. While the redshift cut of MaNGA is apparent, and the MaNGA PSBs have systematically lower stellar masses as a result, we see that the MaNGA PSBs nonetheless span a wide range in stellar mass. Furthermore, when we consider the trends in stellar mass and angular momentum in Figure \ref{fig: lambda_vs_mass}, we see that the PSBs have higher $\lambda_{Re}$ than the ETG sample over a wide range in stellar mass. Thus, the lack of higher stellar mass PSBs at larger distance does not hinder our primary conclusions. When we compare the range of other quantities, namely the stellar age as traced by Dn4000 and ellipticity, a Kolmogorov-Smirnov test shows that we cannot exclude the null hypothesis that the MaNGA PSBs are consistent with the parent sample of SDSS PSBs.

\section{Conclusions}
\label{sec: conclusion}
We trace the evolution of a galaxy as it evolves from the post-starburst phase to the quiescent phase by comparing and contrasting spatially resolved properties of 89 post-starbursts (PSBs) from the MaNGA survey with 260 early-type galaxies (ETGs) from the ATLAS$^{3D}$ project, 1592 MaNGA star-forming galaxies (SFGs), and 916 ETGs from MaNGA. We attempt to glean information about the merger history of these galaxies by studying their inner kinematics and classifying them as slow and fast rotators based on their angular momentum and ellipticity at the effective radius. Our conclusions are as follows:

\begin{itemize}

    \item[(i)] We find that the PSB sample has fewer slow rotators than the ETG samples from both ATLAS$^{3D}$ and MaNGA, with the PSB sample consisting of $\sim 6\%$ (5/89), the ATLAS$^{3D}$ ETG sample having $\sim 14\%$ (36/260), and MaNGA ETGs having $\sim 19.5\%$ (196/916) slow rotators. The SFGs from MaNGA have fewer slow rotators ($\sim3.5\%$, 56/1592) than the PSB and ETG samples, with the median angular momentum of the SFG sample higher than that of the fast-rotator PSB and ETG samples.

    \item[(ii)] We note that the angular momentum is generally lower for ETGs with higher stellar mass, which is not the case for the PSBs or SFGs. This is evident from our finding that $\sim58\%$ of slow-rotator MaNGA ETGs and $\sim58\%$ of slow-rotator ATLAS$^{3D}$ ETGs are in the top 30$^{th}$ percentile of the respective ETG stellar mass distributions, while only $40\%$ slow-rotating MaNGA PSBs and $\sim32\%$ slow-rotating MaNGA SFGs are in the top 30$^{th}$ percentile of the respective PSB and SFG stellar mass distribution.
    
    \item[(iii)] From our analysis of the angular momentum in the context of mass-weighted (MW) and luminosity-weighted (LW) ages, we find that the angular momentum of SFGs and ETGs is more tightly correlated with MW and LW ages as compared to the PSBs. We observe a significant decrease in angular momentum with age for the ETG and SFG samples, though not within the PSB sample.
    
    \item[(iv)] Our analysis of galaxy kinematics with respect to asymmetry does not suggest a clear trend for the PSBs (Spearman correlation: $\rho \approx$ 0.15, p $\approx$ 0.21). Both slow- and fast-rotator ETGs have similarly low values of asymmetry, while the SFG sample displays a correlation of higher asymmetry in the highest-angular-momentum galaxies.
\end{itemize}

This work constrains the possible formation and evolution of ETGs. Roughly half of the ETGs are expected to have gone through a PSB phase after rapidly quenching \citep{2020MNRAS.494..529W}, so if slow-rotator ETGs are created through gas-rich mergers, we should observe a similar or greater fraction of slow rotators in the PSBs. However, we find more slow rotators in the ETG sample as compared to the PSBs. The tighter anticorrelation of angular momentum with the MW and LW age for the SFGs and ETGs, as compared to the PSBs, indicates that angular momentum is more closely coupled to the longer-scale formation history and evolution of the galaxy while it is in the SFG and ETG phases, instead of the time it spends in the green valley region. PSBs already have lower angular momentum than star-forming/disk-like galaxies, and since kinematic evolution takes place over the multi-Gyr lifetime of a galaxy, PSBs must still lose significant angular momentum before evolving to ETGs. This is consistent with the trends in angular momentum with stellar mass, which imply that the slow-rotator ETGs have gained mass and lost angular momentum over many billion years since they last formed stars in situ or experienced a starburst. The long timescale of kinematic evolution is probably why we do not observe significant trends for angular momentum with asymmetry as the asymmetry varies quite rapidly during/after a merger. These implications suggest that the PSBs will undergo multiple major/minor dry merger events as they evolve to the ETG phase while shedding off their angular momentum. This scenario is consistent with the findings of simulations presented in \citet{2014MNRAS.444.3357N}, where the slow-rotator ETGs tended to have many dry mergers while being quiescent for a long time.

\section*{Acknowledgements}
\label{sec: acknowledgements}
We thank the referee for their comments, which have improved this manuscript. We thank Akshat Tripathi for useful discussions. This work was made possible through the Preparing for Astrophysics with LSST Program, supported by the Heising-Simons Foundation and managed by Las Cumbres Observatory.

Funding for the Sloan Digital Sky Survey IV has been provided by the Alfred P. Sloan Foundation, the U.S. Department of Energy Office of Science, and the Participating Institutions. SDSS-IV acknowledges support and resources from the Center for High Performance Computing at the University of Utah. The SDSS website is www.sdss4.org.

SDSS-IV is managed by the Astrophysical Research Consortium for the Participating Institutions of the SDSS Collaboration including the Brazilian Participation Group, the Carnegie Institution for Science, Carnegie Mellon University, Center for Astrophysics | Harvard \& Smithsonian, the Chilean Participation Group, the French Participation Group, Instituto de Astrof\'isica de Canarias, The Johns Hopkins University, Kavli Institute for the Physics and Mathematics of the Universe (IPMU) / University of Tokyo, the Korean Participation Group, Lawrence Berkeley National Laboratory, Leibniz Institut f\"ur Astrophysik Potsdam (AIP),  Max-Planck-Institut f\"ur Astronomie (MPIA Heidelberg), Max-Planck-Institut f\"ur Astrophysik (MPA Garching), Max-Planck-Institut f\"ur Extraterrestrische Physik (MPE), National Astronomical Observatories of China, New Mexico State University, New York University, University of Notre Dame, Observat\'ario Nacional / MCTI, The Ohio State University, Pennsylvania State University, Shanghai Astronomical Observatory, United Kingdom Participation Group, Universidad Nacional Aut\'onoma de M\'exico, University of Arizona, University of Colorado Boulder, University of Oxford, University of Portsmouth, University of Utah, University of Virginia, University of Washington, University of Wisconsin, Vanderbilt University, and Yale University.

\software{Astropy \citep{2013A&A...558A..33A, 2018AJ....156..123A, 2022ApJ...935..167A}, Matplotlib \citep{4160265}, NumPy \citep{2020Natur.585..357H}, SDSS-Marvin \citep{2019AJ....158...74C}, Pandas \citep{pandas}, Seaborn \citep{Waskom2021}, SciPy \citep{2020SciPy-NMeth}}

\bibliography{ref}
\bibliographystyle{aasjournal}

\appendix

\section{Comparison of Photometry}
\label{sec: photometry_comparison}
In order to analyze the kinematics within 1 $Re$, we need to know the shape and size of the galaxy. We also need the Sérsic index to estimate the impact of beam smearing on the photometry. We compare the photometric properties from the NSA catalog with the ones given in Table 3 of \citet{2011ApJS..196...11S}. \citet{2011ApJS..196...11S} performed bulge+disk decompositions in the $g$ and $r$ bands by convolving the PSF in 2D for 1,123,718 galaxies from the Legacy portion of SDSS Data Release 7. They use three fitting models: a pure Sérsic model, an n$_b$ = 4 bulge + disk model, and a Sérsic (free n$_b$) bulge + disk model. We chose the data calculated using the pure Sérsic model as NSA also used a Sérsic fit, making the two sets of photometric properties comparable. We compare the galaxy Sérsic index ({\tt ng}), the galaxy position angle ({\tt phi}), galaxy ellipticity ({\tt e}), and the galaxy half-light radius in the $r$ band ({\tt Rhlr}). 

\begin{figure*}
\epsscale{1.2}
\plotone{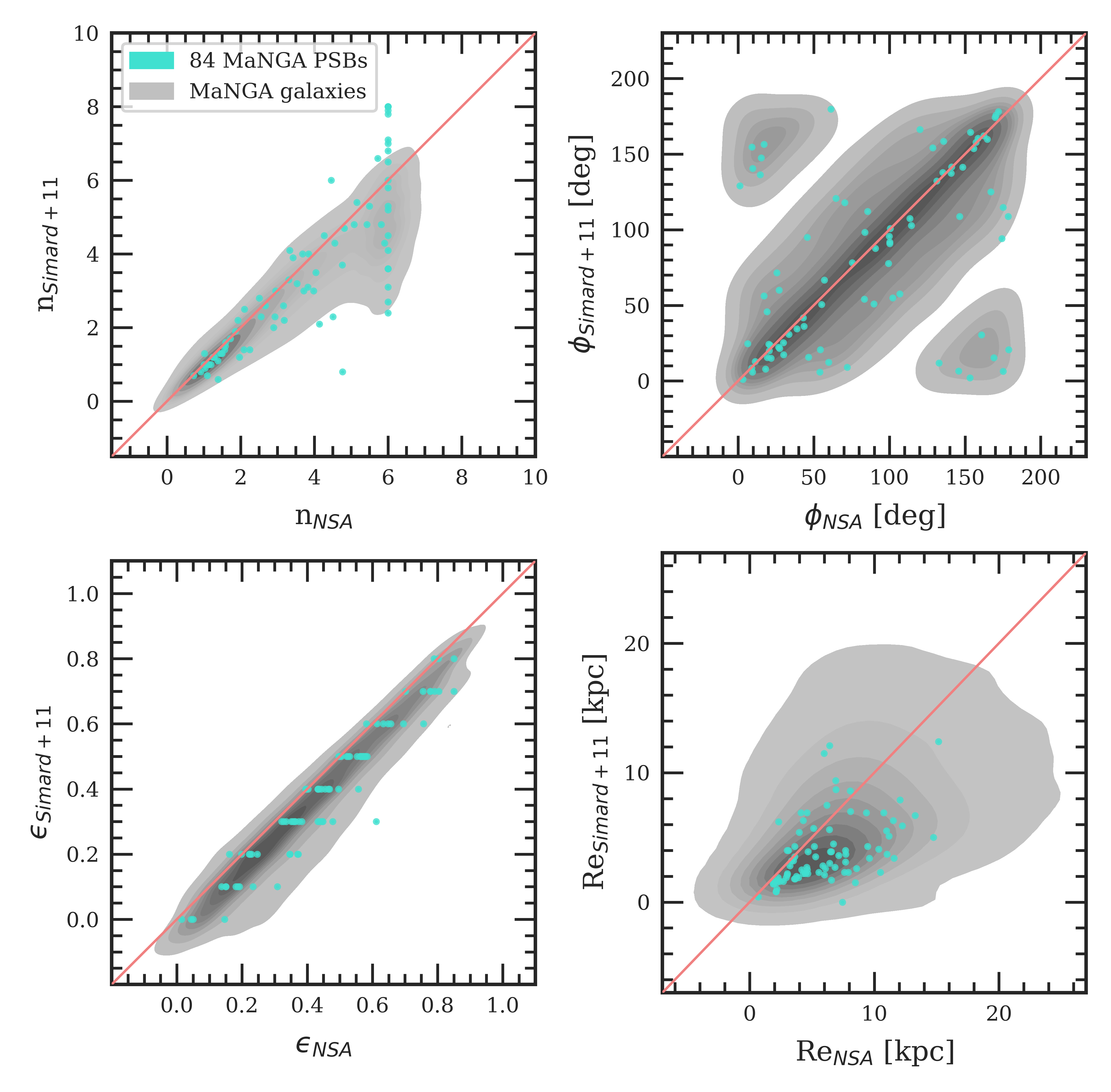}
\caption{The four panels compare the Sérsic index (top left), photometric angle (top right), ellipticity (bottom left), and half-light radius (bottom right) of all the galaxies from the NSA catalog with the properties given in Table 3 of \citet{2011ApJS..196...11S}, all of which were determined using a pure Sérsic model. The gray shaded contour region corresponds to all MaNGA galaxies cross-matched with the \citet{2011ApJS..196...11S} data, while the blue points represent the 84 MaNGA PSBs. We use the red 1:1 line to check the consistency of the scatter across the PSB sample with all the galaxies in the two catalogs. We find that the PSBs are similarly distributed as the total galaxy sample, though we observe significant scatter, and some systematic offsets. In Figure \ref{fig: simard_pymorph_vs_nsa}, we test the impact of these offsets on our measurement of $\lambda_{Re}$.
\label{fig: simard_vs_nsa}}
\end{figure*}

To test the consistency of the photometry across all the galaxies, and not just the PSBs, we plot the photometric properties from the NSA catalog against the respective properties from \citet{2011ApJS..196...11S}, as shown in Figure \ref{fig: simard_vs_nsa}. The gray contours represent all the galaxies obtained by cross-matching MaNGA with \citet{2011ApJS..196...11S} data irrespective of galaxy type, and the blue points represent the 84 PSBs with photometric data available in \citet{2011ApJS..196...11S}. We observe that the scatter in the photometric properties of the PSBs around the red 1:1 line is quite consistent with the overall scatter of all the galaxies for all the properties, though there are some systematic offsets in photometric angle and effective radius.

\begin{figure*}
\epsscale{1.2}
\plotone{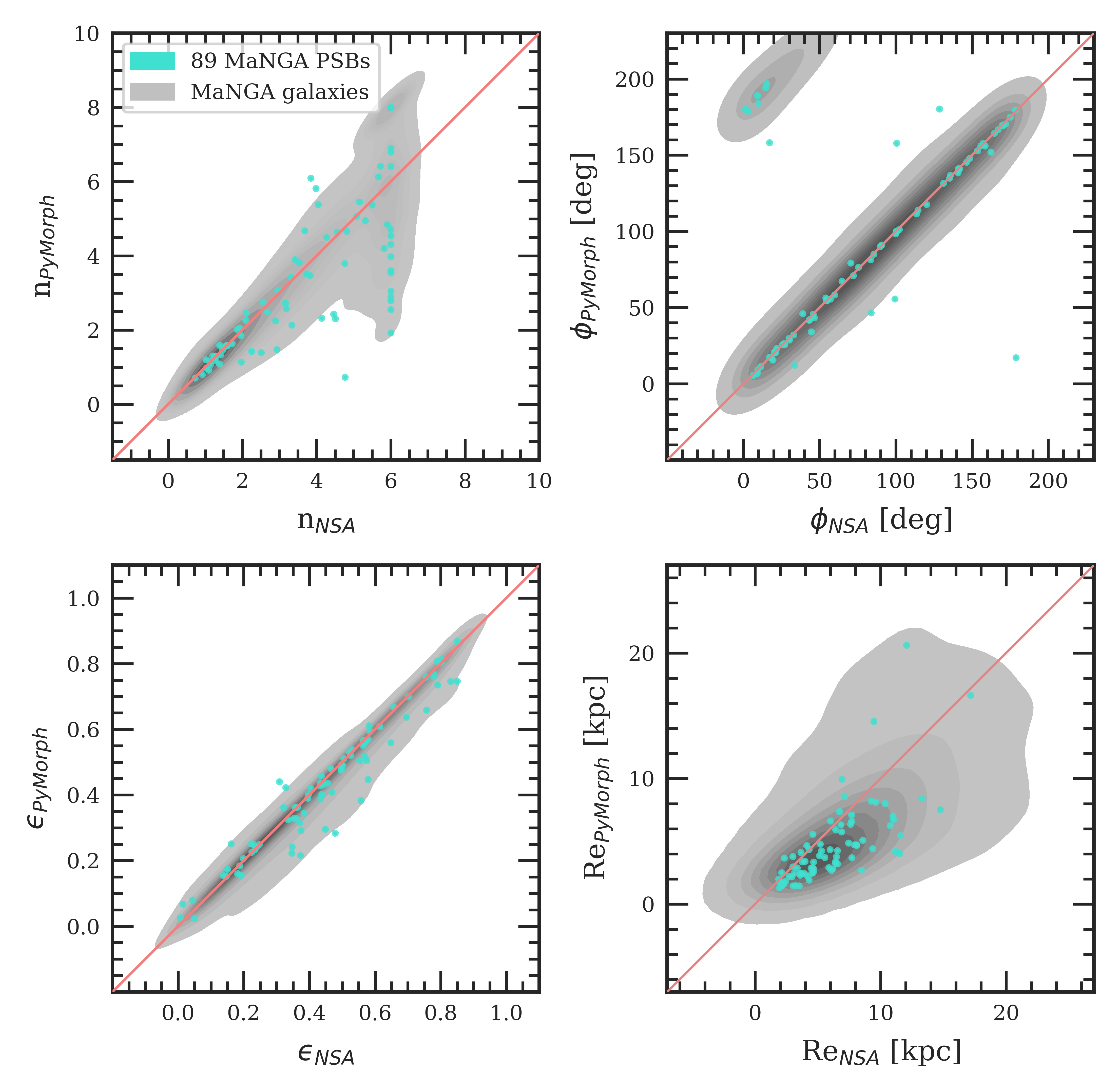}
\caption{The four panels compare the Sérsic index (top left), photometric angle (top right), ellipticity (bottom left), and half-light radius (bottom right) of all the galaxies from the NSA catalog with the corresponding pure Sérsic model properties provided in MaNGA PyMorph DR17 photometric catalogue. The gray shaded contour region corresponds to all MaNGA galaxies cross-matched with the PyMorph catalogue, while the blue points represent the 89 PSBs. We use the red 1:1 line to check the consistency of the scatter across the PSB sample with all the galaxies in the two catalogs. We find that the PSBs are similarly distributed as the total galaxy sample, though we observe significant scatter, and some systematic offsets. In Figure \ref{fig: simard_pymorph_vs_nsa}, we test the impact of these offsets on our measurement of $\lambda_{Re}$.
\label{fig: pymorph_vs_nsa}}
\end{figure*}

\begin{figure*}
\begin{center}
\includegraphics[width=0.49\textwidth]{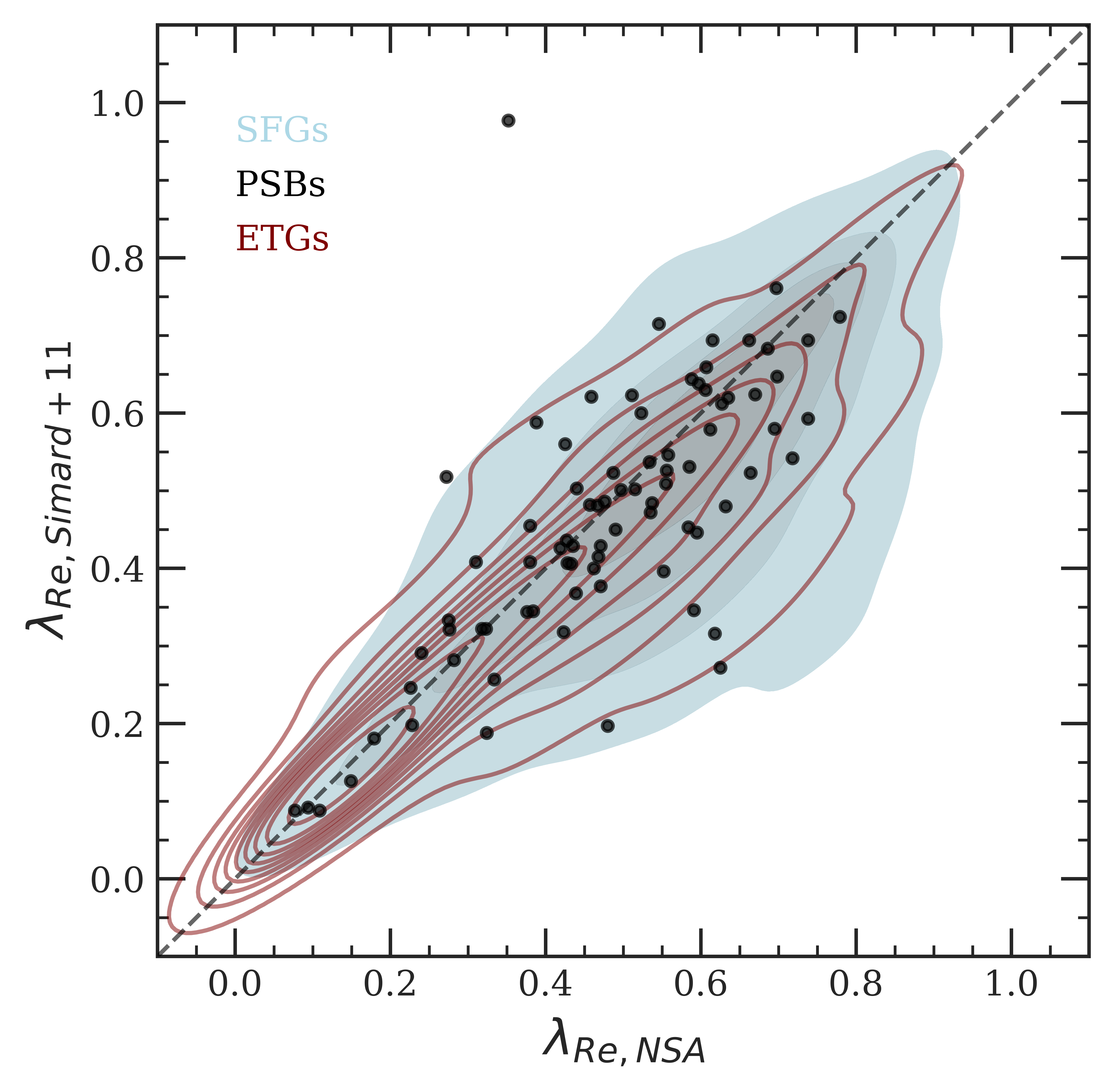}
\includegraphics[width=0.49\textwidth]{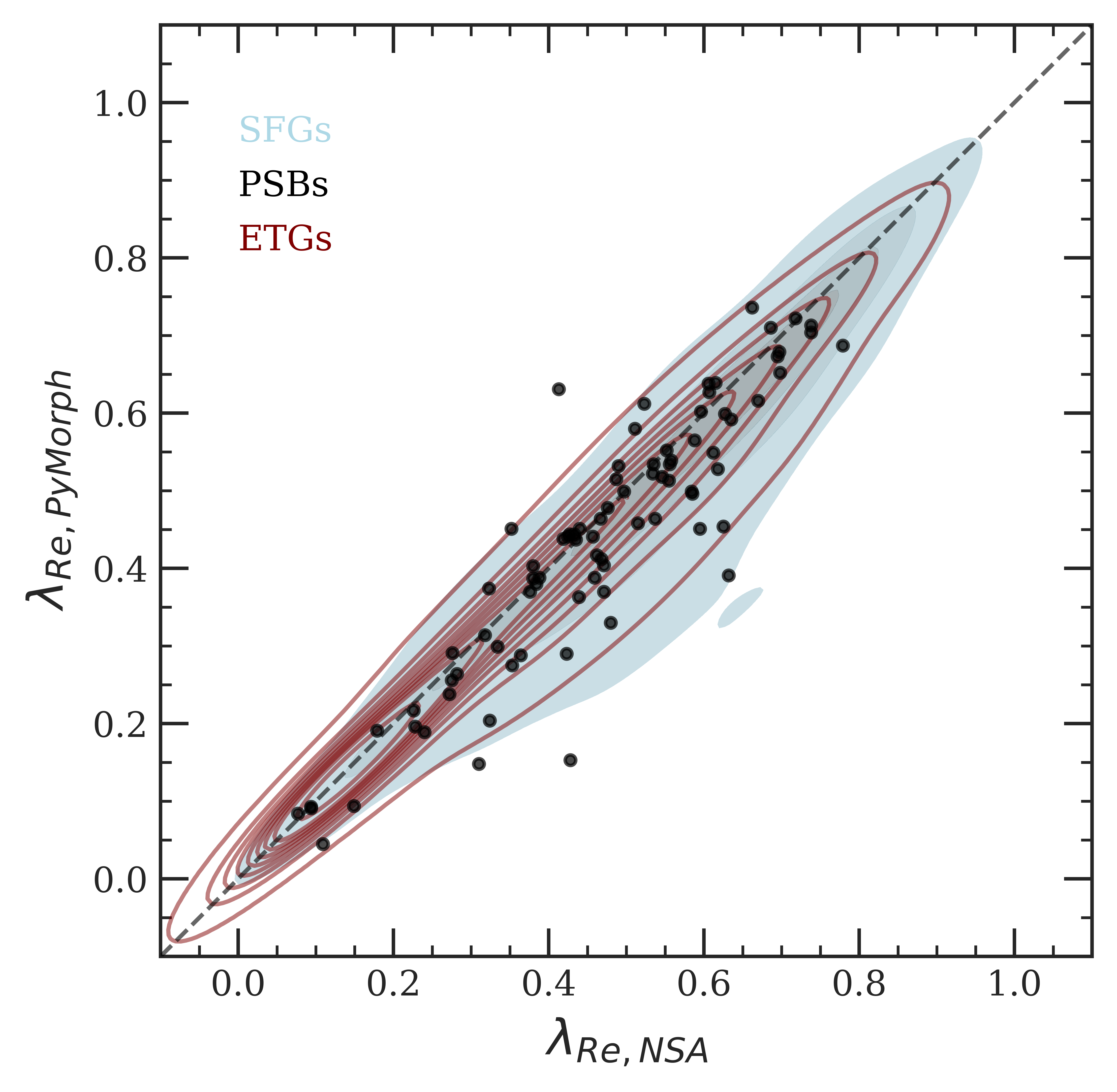}
\end{center}
\caption{We test whether our results are consistent with respect to photometry by comparing the results from the NSA photometric properties with that of MaNGA PyMorph DR17 photometric catalogue and \citet{2011ApJS..196...11S} photometric data. Left: comparison of the angular momentum obtained using photometry from the NSA catalog and \citet{2011ApJS..196...11S}, in conjunction with MaNGA DR17 maps for 81 PSBs (black points), 1512 SFGs (blue contours) and 831 ETGs (red contours). The $\lambda_{Re}$ values determined using \citet{2011ApJS..196...11S} data tend to be lower in comparison to the predictions from NSA photometry, and there is significant scatter about the 1:1 black line, especially for higher values of $\lambda_{Re}$. Using data from \citet{2011ApJS..196...11S}, we find $\sim 5\%$ slow rotators in the SFGs, $\sim 7.4\%$ slow rotators in the PSBs and $\sim 14.8\%$ slow rotators in the ETG sample. Right: we compare NSA with PyMorph photometry, while using MaNGA DR17 maps in our calculation of $\lambda_{Re}$ for 83 PSBs (black points), 1502 SFGs (blue contours) and 853 ETGs (red contours). We observe even less scatter about the 1:1 black line as compared to the left panel, and PyMorph predicts smaller values of $\lambda_{Re}$ as compared to NSA as well. Using PyMorph photometric properties, we find $\sim 5.7\%$ slow rotators in the SFGs, $\sim 8.4\%$ slow rotators in the PSBs and $\sim 19.1\%$ slow rotators in the ETG sample. Irrespective of the catalog being used, the PSBs have fewer slow rotators than the ETGs and tend to rotate faster than the SFGs. This confirms that the choice of photometry would not have a significant impact on our overall results.
}
\label{fig: simard_pymorph_vs_nsa}
\end{figure*}

To further test our choice to use the NSA photometry, we compare the NSA catalog to the MaNGA PyMorph DR17 photometric catalogue. PyMorph provides photometric properties determined by fitting Sérsic and Sérsic+Exponential models to the 2D surface brightness profiles of galaxies in the MaNGA DR17 release \citep{2013MNRAS.433.1344M, 2017MNRAS.467..490F, 2017MNRAS.468.2569B, 10.1093/mnras/stab3089}. PyMorph was rerun to incorporate the SDSS DR14 images, improved bulge-to-disk decomposition, and all the fits were re-fit if deemed necessary \citep{10.1093/mnras/sty3135, 10.1093/mnras/stab3089}. Since NSA performed a pure Sérsic fit, we selected the PyMorph photometry obtained from fitting the Sérsic model. We compare the Sérsic index ({\tt N\_S}), photometric angle (90 - {\tt PA\_S} - {\tt SPA}), ellipticity (1-{\tt BA\_S}), and effective radius (obtained from the semi-major axis, {\tt A\_hl\_S}) in the $r$ band from PyMorph runs with the corresponding properties in the $r$ band from the NSA catalog, as shown in Figure~\ref{fig: pymorph_vs_nsa}. The gray contours represent all the galaxies obtained by cross-matching MaNGA with the PyMorph catalogue irrespective of phase, and the blue points represent the 89 PSBs across the two catalogs. Similar to the comparison between the MaNGA and \citet{2011ApJS..196...11S} data, we observe similarities in the PSB sample with the overall galaxy sample, as well as systematic offsets and scatter. The impact of this offset and scatter in the photometric properties on $\lambda_{Re}$ is further explored in Figure~\ref{fig: simard_pymorph_vs_nsa}.

The scatter observed in photometric properties between these catalogs may mean that the choice of catalog influences our measurement of specific angular momentum. The $\lambda_{Re}$ obtained using the three photometric catalogs, MaNGA DR17 data, and the method we describe above are plotted against each other in Figure~\ref{fig: simard_pymorph_vs_nsa}. The Sérsic indices were modified as explained in Section \ref{sec: calculation} to ensure that they are in the range $0.5-6.5$ for 11/84 PSBs from the \citet{2011ApJS..196...11S} sample and 15/89 PSBs from the PyMorph catalogue. We cannot accurately measure $\lambda_{Re}$ for 3/84 PSBs as their semimajor axis from \citet{2011ApJS..196...11S} data is smaller than their $\sigma_{PSF}$, so we only include 81/84 PSBs in Figure \ref{fig: simard_pymorph_vs_nsa}(a). Similarly, Figure \ref{fig: simard_pymorph_vs_nsa}(b) consists of only 83/89 PSBs due to missing $\epsilon_{Re}$ measurements for the excluded galaxies in PyMorph photometry. The $\lambda_{Re}$ values predicted using PyMorph agree better with the predictions from NSA photometric data as compared to the \citet{2011ApJS..196...11S} data.  We find that, on average, \citet{2011ApJS..196...11S} and PyMorph catalogs predict lower $\lambda_{Re}$ as compared to NSA. Using \citet{2011ApJS..196...11S} photometry, we find $\sim 5\%$ slow rotators in the SFGs, $\sim 7.4\%$ slow rotators in the PSBs, and $\sim 14.8\%$ slow rotators in the ETGs. Similarly, we find $\sim 5.7\%$ slow rotators in the SFGs, $\sim 8.4\%$ slow rotators in the PSBs, and $\sim 19.1\%$ slow rotators in the ETG sample using PyMorph photometry. NSA photometry predicts the least number of slow rotators for the SFGs and PSBs as compared to the \citet{2011ApJS..196...11S} and PyMorph catalogs. However, the PSBs have fewer slow rotators than the ETGs and rotate faster as compared to the SFGs irrespective of the photometry utilized, which confirms that choosing another photometric catalog would not affect our qualitative results and conclusions drastically. Thus, the NSA catalog is a suitable choice for measuring $\lambda_{Re}$ in this work as described in Section \ref{sec: calculation}. 

\begin{figure*}
\begin{center}
\includegraphics[width=0.49\textwidth]{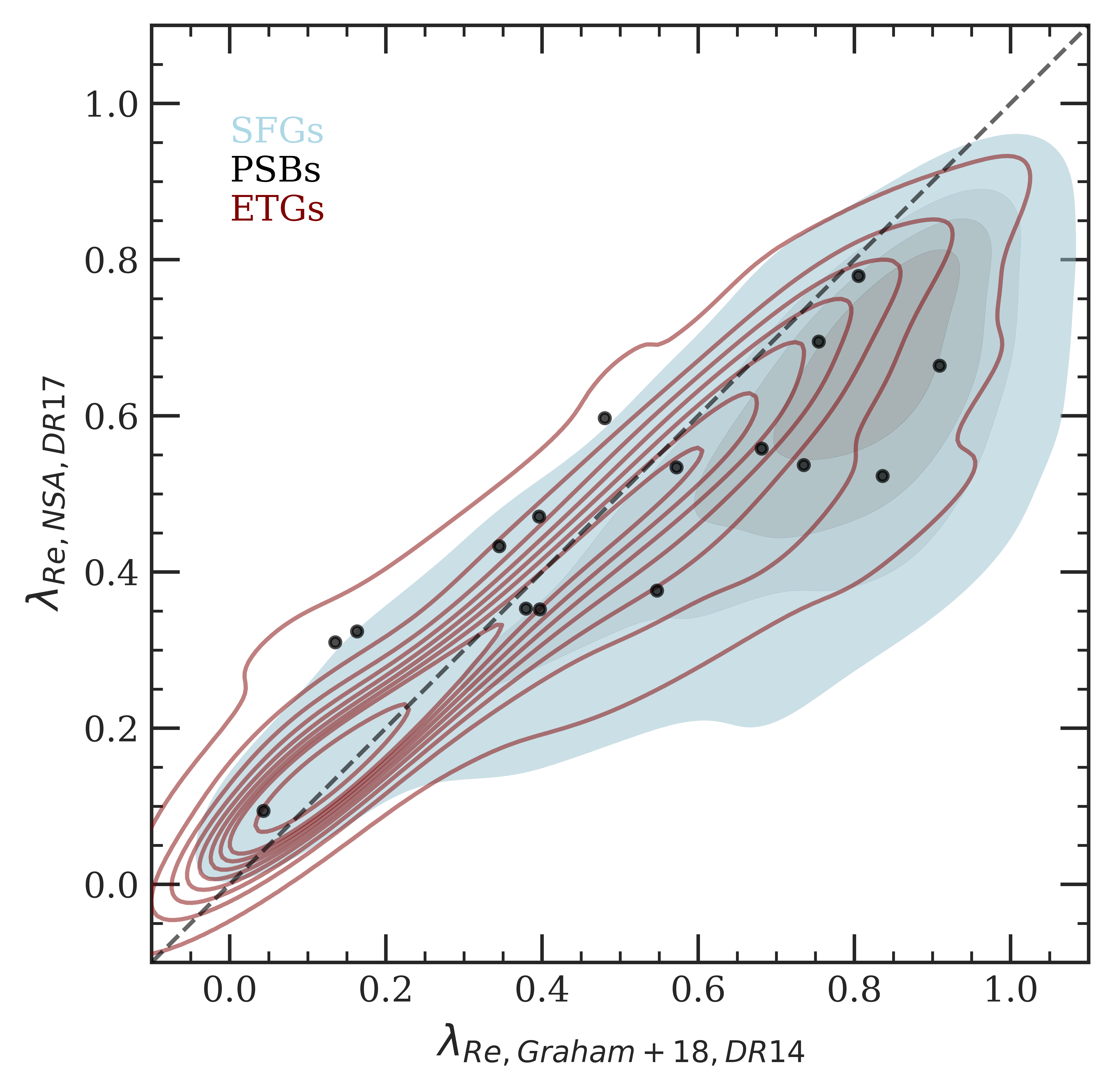}
\includegraphics[width=0.49\textwidth]{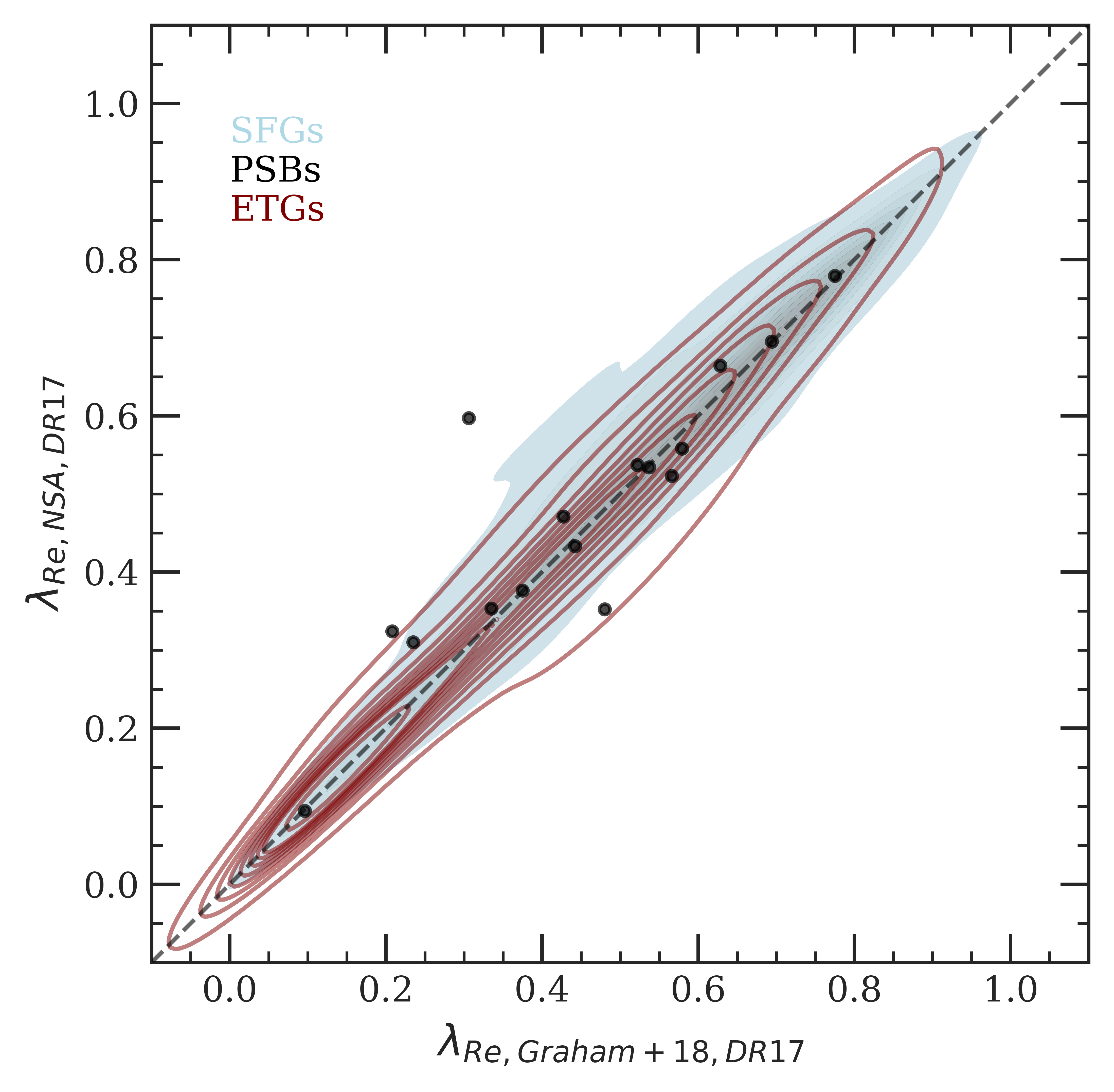}
\end{center}
\caption{We check for the impact of the MaNGA data release used on our results by drawing comparisons between DR14 and DR17 data using photometry from \citet{2018MNRAS.477.4711G} and NSA. Left: we compare the $\lambda_{Re}$ obtained from photometric properties determined by \citet{2018MNRAS.477.4711G} and DR14 with the values obtained using DR17 in conjunction with NSA photometry. The black points are the 16 MaNGA PSBs, the blue contours are the 1571 SFGs and the red contours are the 914 ETGs that are present in both the photometric catalogs and MaNGA data releases. We find that the $\lambda_{Re}$ values are scattered around the 1:1 (black dashed) line, with the scatter increasing for higher values of $\lambda_{Re}$ for all three classes. Right: to analyze how the difference in the data release affects our results, we set the data release to DR17 and calculate $\lambda_{Re}$ using \citet{2018MNRAS.477.4711G} and NSA photometry respectively for 16 PSBs, 1569 SFGs and 914 ETGs present in both of the photometric catalogs. We observe a significant reduction in scatter around the 1:1 line for all three classes, owing to the most recent data release being utilized. As evident from the two panels, the MaNGA data release used for obtaining the kinematic spectra has a non-negligible impact on the kinematic analysis, which contains key implications for the overall kinematic evolution of the galaxy.}
\label{fig: lambda_vs_lambda}
\end{figure*}

We perform a similar test on the different data releases of MaNGA to compare directly with the measurements from \citet{2018MNRAS.477.4711G} and check for any systematic differences. We compare MaNGA DR14 and DR17 data along with photometric properties from \citet{2018MNRAS.477.4711G} and the NSA catalog while uniformly using the function given by \citet{2018MNRAS.477.4711G} for calculation of $\lambda_{Re}$ in Figure \ref{fig: lambda_vs_lambda}. In Figure \ref{fig: lambda_vs_lambda}(a), we compare the $\lambda_{Re}$ from \citet{2018MNRAS.477.4711G} (which uses MaNGA DR14 and photometric properties derived in that work) and the $\lambda_{Re}$ we recalculate using DR17 and the NSA photometry. As evident from Figure \ref{fig: lambda_vs_lambda}(a), the SFGs, PSBs and ETGs seem to scatter around the 1:1 line (black dashed line). We also observe systematic shifts that differ between samples. The scatter in both the samples increases with $\lambda_{Re}$. In Figure \ref{fig: lambda_vs_lambda}(b), we compare the $\lambda_{Re}$ calculated using photometric properties derived by \citet{2018MNRAS.477.4711G} with NSA photometry while setting the data release to DR17 for both. Fixing the data release to DR17 and only varying the photometric properties results in a significant reduction in the scatter in $\lambda_{Re}$ for the SFGs, PSBs and ETGs. This shows that the data release being utilized in the calculation for $\lambda_{Re}$ has a significant impact on our results, more so than the photometric parameters used.

Because of the improvements in DR17 that we discuss in Section \ref{sec: MaNGA}, we use our rereduced $\lambda_{Re}$ values (the y-axis of Figures \ref{fig: lambda_vs_lambda}(a) and (b)) for the SFG and ETG samples in this manuscript. The method for calculating $\lambda_{Re}$ is thus the same for both the PSB and comparison samples.

\end{document}